\documentclass[aps,preprint,showpacs,showkeys,nofootinbib]{revtex4-1}
\pdfoutput=1
\RequirePackage[T1]{fontenc}

\usepackage{graphicx}
\usepackage{color}
\usepackage{mathtools} 
\usepackage{amsmath}
\usepackage{caption}
\usepackage{subcaption}
\usepackage{float}
\usepackage{color}
\usepackage{multirow}
\usepackage{bbold}
\usepackage{array}
\usepackage{tabularx}
\usepackage{hyperref}
\usepackage{calc}
\usepackage[utf8x]{inputenc}
\definecolor{My_red}{cmyk}{0.00,1.00,1.00,0.20}

\usepackage{slashed, latexsym, verbatim, cancel}
\usepackage{pstricks}
\usepackage{bbm}

\newcommand\inv[1]{#1\raisebox{1.4ex}{$\scriptscriptstyle-\!1$}}

\newcommand{\bmat}{\left(\begin{array}}
\newcommand{\emat}{\end{array}\right)}
\newcommand{\beq}{\begin{equation}}
\newcommand{\eeq}{\end{equation}}

\newcommand{\VEV}[1]{\langle  #1 \rangle}
\newcommand{\mfrac}[2]{\frac{ \mbox{$#1$} }{ \mbox{$#2$} }}

\def\bwt{\begin{widetext}}
\def\ewt{\end{widetext}}
\def\be{\begin{equation}}
\def\ee{\end{equation}}
\def\bea{\begin{eqnarray}}
\def\eea{\end{eqnarray}}
\def\bean{\begin{eqnarray*}}
\def\eean{\end{eqnarray*}}
\def\bary{\begin{array}}
\def\eary{\end{array}}
\def\bit{\begin{itemize}}
\def\eit{\end{itemize}}

\def\su5u1{SU(5) \times U(1)}
\def\fsu5u1{SU(5) \times U(1)'}
\def\so10{SO(10)}
\def\sq20{SO(10) \times SO(10)}

\def\nn{\nonumber}

\def\bwt{\begin{widetext}}
\def\ewt{\end{widetext}}
\def\be{\begin{equation}}
\def\ee{\end{equation}}
\def\bea{\begin{eqnarray}}
\def\eea{\end{eqnarray}}
\def\bean{\begin{eqnarray*}}
\def\eean{\end{eqnarray*}}
\def\bary{\begin{array}}
\def\eary{\end{array}}
\def\bit{\begin{itemize}}
\def\eit{\end{itemize}}

\def\su5u1{SU(5) \times U(1)}
\def\fsu5u1{SU(5) \times U(1)'}
\def\so10{SO(10)}
\def\sq20{SO(10) \times SO(10)}

\usepackage{amssymb}

\begin{document}
\title{Non-standard signatures of vector-like quarks in a leptophobic $221$ model.}

\author{Kasinath Das}
\email{kasinath.das91@gmail.res.in}

\affiliation{Regional Centre for Accelerator-based Particle Physics, Harish-Chandra Research Institute, HBNI, Chhatnag Road, Jhusi, Allahabad 211019, India}

\author{Tanmoy Mondal}
\email{tanmoymondal@hri.res.in}

\affiliation{Regional Centre for Accelerator-based Particle Physics, Harish-Chandra Research Institute, HBNI, Chhatnag Road, Jhusi, Allahabad 211019, India}

\author{Santosh Kumar Rai}
\email{skrai@hri.res.in}

\affiliation{Regional Centre for Accelerator-based Particle Physics, Harish-Chandra Research Institute, HBNI, Chhatnag Road, Jhusi, Allahabad 211019, India}

\preprint{HRI-RECAPP-2018-007}
\date{\today}

\begin{abstract}
We consider  vector-like quarks in a leptophobic 221 model characterized by the gauge group $SU(2)_L\times SU(2)_2 \times U(1)_X$, 
where the $SU(2)_2$ is leptophobic in nature. We discuss about the pattern of mixing between Standard Model quarks and 
vector-like quarks and how we prevent tree level flavour-changing interactions in the model. The model also predicts tauphilic scalars 
decaying mostly to tau leptons. We consider a typical signal of the model in the form of pair production of top-type vector-like quarks 
which decays to the tauphilic scalars and a third generation quark. We analyze the resulting final state signal for the 13 TeV LHC, 
containing $\geq 3j(1b) \, + \geq 2\tau \,  + \geq 1l$ and discuss the discovery prospects of such vector-like quarks with non-standard 
decay modes.
\end{abstract}

\keywords{Extended Gauge Models, Vector-like Quarks, Flavour Changing Neutral Currents}

\maketitle
\section{Introduction}
As the Large Hadron Collider (LHC) churns out more and more data and gets ready for an energy upgrade, lack of new physics 
signal at the high energy frontier only makes us more intrigued with what picture of beyond Standard Model (SM) could eventually 
emerge. The SM itself highlights the great success of gauge theory and a natural extension would be in the form of additional gauge 
symmetries with new matter fields.  We know that all the three generations of matter fields in the SM are chiral in nature. The 
possibility of a fourth generation of chiral fermions, especially quarks has been excluded by the Higgs signal strength measurements 
along with the electroweak precision data \cite{Eberhardt:2012gv}. However the possibility of having vector-like quarks (VLQ) whose 
left- and right-chiral components  transform 
in the same way under the SM gauge group still exists and are being searched for at the LHC.

The collider signatures of a VLQ depends on its possible decay modes. The existing searches for VLQs are under the assumption that 
they decay to a SM boson and a SM quark. For example searches on top-like VLQ with electric charge $Q=+\frac{2}{3}$ assume that
it decays to $Z\,t$, $W^+b$ and  $h\,t$ and for a bottom-like VLQ with electric charge $Q=-\frac{1}{3}$, the decay modes are $Z\,b$, $W^-t$ and $h\,b$. 
Assuming strong pair production, both ATLAS and CMS collaborations have obtained different lower limits on the masses of third 
generation VLQs for different branching ratio hypotheses \cite{Sirunyan:2018omb, Sirunyan:2017pks, Aaboud:2018saj, Aaboud:2018uek, Aaboud:2018xuw, Aaboud:2017zfn}.
The most stringent lower bound on the mass of a top-like VLQ obtained by using the LHC Run 2 data is 1.3 TeV given by 
CMS \cite{Sirunyan:2018omb} and 1.43 TeV given by ATLAS \cite{Aaboud:2018xuw} while the most stringent lower bound on the mass of 
a bottom-like VLQ is 1.24 TeV given by CMS \cite{Sirunyan:2018omb} and 1.35 TeV given by ATLAS \cite{Aaboud:2018uek}. Since the single production of VLQs depend 
on the mixing between VLQs and SM quarks, based on the searches for single-production of VLQs both CMS and ATLAS collaborations have given 
exclusion limits for the product of production cross section and branching fraction for different mass values \cite{Sirunyan:2018fjh, Sirunyan:2017ynj, ATLAS:2016ovj}. 
Extensive phenomenological studies on VLQs in standard decay scenarios exist in literature \cite{AguilarSaavedra:2005pv, AguilarSaavedra:2009es, Cacciapaglia:2010vn,
Okada:2012gy, DeSimone:2012fs, Aguilar-Saavedra:2013qpa, Ellis:2014dza}.

Exotic fermions are a necessary ingredient in some gauge extended models for anomaly cancellation, e.g., exotic quarks in leptophobic 221 model \cite{Hsieh:2010zr, Das:2015ysz},
exotic leptons in hadrophobic 221 model \cite{Hsieh:2010zr} and exotic fermions in almost all $U(1)^\prime$ extensions \cite{Langacker:2008yv}.
These exotic fermions become vector-like once the full symmetry group breaks down to the SM gauge group. These fermions are quasi-chiral in nature,
i.e., they are vector-like under the SM gauge group but chiral under the extra gauge group \cite{Langacker:2008yv}.

VLQs in gauge extended models can have interesting collider signatures because the rich spectrum of the model opens up non-standard decay modes for the VLQs.
In this work we have considered the collider signatures of certain non-standard decay modes of top-like VLQ in a leptophobic 221 model
characterized by the gauge group $SU(2)_1\times SU(2)_2\times U(1)_X$. Because of the presence of the non-standard decay modes
the existing constraints on the masses of VLQs will get significantly relaxed and the VLQ can lie at the sub-TeV scale.
Phenomenological studies of VLQs having non-standard decay modes exist in literature for various non-minimal extensions of SM \cite{Kearney:2013oia, Kearney:2013cca, 
Karabacak:2014nca, Serra:2015xfa, Anandakrishnan:2015yfa, Banerjee:2016wls, Arhrib:2016rlj, Dobrescu:2016pda, Aguilar-Saavedra:2017giu, Chala:2017xgc, Moretti:2017qby, Chala:2018qdf, Bizot:2018tds, Kim:2018mks}. 
Collider signatures of vector-like quarks in $U(1)$ gauge extensions have been considered in \cite{Grossmann:2010wm, Joglekar:2016yap, Das:2017fjf}.

In general, the mixing between SM quarks and VLQs which play an important role in the phenomenology of VLQs also generate tree level 
flavour-changing neutral current (FCNC) interactions of SM quarks with the $Z$ boson. 
Several studies on the mixing of VLQs with SM quarks are available in the literature \cite{AguilarSaavedra:2002kr, Cacciapaglia:2011fx, Aguilar-Saavedra:2013qpa, Aguilar-Saavedra:2013wba, Alok:2014yua, Alok:2015iha, Chen:2017hak} which take into account constraints from flavour 
physics and electroweak precision measurements .
Mixings are strongly constrained from FCNC processes. Even in the presence of mixings with VLQs the possibility of avoiding tree level 
FCNC interactions is possible for judiciously chosen mixing patterns \cite{Langacker:1988ur}. 
We discuss such a mixing pattern for the leptophobic 221 model which avoids tree level flavour-changing interactions with the $Z$ and $Z^\prime$. For general mixing scenarios between VLQs and SM quarks, all the neutral scalars present in the model have flavour-changing 
interactions, thus making it difficult to get a 125 GeV neutral Higgs and simultaneously satisfying constraints from FCNC processes.
We show that it is possible to avoid flavour-changing interactions for certain neutral scalars (non-FCNH scalars), which can then lie at 
sub-TeV scale. In this work we study the collider signatures of the third generation top-type VLQ decaying to a final state with one of 
these non-FCNH scalars (other than the 125 GeV Higgs) and a third generation SM quark and the scalars then decay dominantly to 
tau leptons.
 
The paper is organized as follows. In Section \ref{section:model} we discuss our model. In Section \ref{section:Int_With_Gauge_Bosons} we discuss about the interaction of the VLQs with the SM gauge bosons. In Section \ref{section:FCNC} we discuss the 
pattern of mixing between the SM quarks and the VLQs for which FCNC interactions of SM quarks with $Z$-boson and the FCNH 
interactions with the SM-Higgs boson is zero at the tree level. In Section \ref{section:phenomenology} we discuss the possible 
phenomenology of VLQs in the model and explore the possible collider signatures in Section.\ref{section:collider}. Finally we conclude 
and summarize in Section \ref{section:conclusion}.

\section{The Model}\label{section:model}
The $SU(2)$ extensions of the SM characterized by the gauge group $SU(3)_C \times SU(2)_1 \times SU(2)_2 \times U(1)_X$ are generally called 221 models in the literature. 
Depending on the way the SM fermions transform under the gauge groups, different versions of 221 models are possible, {\it viz.}
leptophobic, hadrophobic and left-right symmetric etc. The model is called leptophobic when the SM right-chiral leptons are singlets 
under $SU(2)_2$, hadrophobic when the right-chiral SM quarks are singlets under $SU(2)_2$ and left-right symmetric when both the 
SM right-chiral quarks and the right-chiral
leptons (with the addition of right-chiral neutrinos to the model) form doublets under $SU(2)_2$. The most popular among all 221 models is the left-right symmetric model 
\cite{Mohapatra:1974gc, Mohapatra:1974hk, Senjanovic:1975rk, Senjanovic:1978ev, Mohapatra:1979ia, Mohapatra:1980yp}. 
A general classification of all 221 models has been done in \cite{Hsieh:2010zr} based on two types of  symmetry breaking patterns of the gauge
group. The two patterns are as follows :
\begin{itemize}
 \item Type-I : $SU(2)_1$ is identified with $SU(2)_L$ of SM. The first stage of symmetry breaking is $SU(2)_2 \times U(1)_X \rightarrow U(1)_Y$. 
 The second stage of symmetry breaking is $SU(2)_L \times U(1)_Y \rightarrow U(1)_{em}$.
  \item Type-II : $U(1)_X$ is identified with $U(1)_Y$  of SM. The first stage of symmetry breaking is $SU(2)_1 \times SU(2)_2 \rightarrow SU(2)_L$. 
 The second stage of symmetry breaking is $SU(2)_L \times U(1)_Y \rightarrow U(1)_{em}$.
\end{itemize}
Certain type of 221 models need exotic fermions for the cancellation of anomalies. For example, the leptophobic 221 model needs 
exotic quarks and the hadrophobic 221 model requires exotic leptons. These exotic fermions become vector-like after the breaking of the 
full symmetry group down to SM gauge group.

In this work we consider the leptophobic 221 model, which follows the type-I symmetry breaking pattern described above and the SM leptons are singlets under the 
$SU(2)_2$ gauge group\footnote{The model has been considered previously in \cite{Das:2015ysz} to explain the reported excess for a 
narrow width resonance around 2 TeV in the $WZ$, $WW$, and $ZZ$ channel by the ATLAS collaboration \cite{Aad:2015owa} using the 20.3 fb$^{-1}$ of data of 8 TeV LHC.}. We will denote $SU(2)_1$ as $SU(2)_L$ throughout the article.
The scalar sector of the model contains two scalar doublets represented by $H_1$ and $H_2$, and a bi-doublet  scalar represented by $\Phi$.
The symmetry breaking of the full gauge group to $U(1)_{em}$ by the scalars occurs in two stages :
\begin{align}
\text{Stage I} &: \, \, \, \, \, SU(2)_2 \times U(1)_X \xrightarrow[]{\langle H_2 \rangle} U(1)_Y, \nn \\
\text{Stage II} &: \, \, \, \, \, SU(2)_L \times U(1)_Y \xrightarrow[]{\, \, \,  \langle \Phi \rangle, \, \langle H_1 \rangle   \,  \, }  U(1)_{em}.
\end{align}
The first stage of symmetry breaking of the full gauge group to the SM gauge group ($SU(2)_L \times U(1)_Y$) is achieved by the vacuum expectation value (VEV) of $H_2$, 
which is a doublet under $SU(2)_2$. 
The subsequent second stage breaking of the SM gauge group to the $U(1)_{em}$ gauge group occurs, once $\Phi$, a bi-doublet under $SU(2)_L\times SU(2)_2$ or $H_1$,
a doublet under $SU(2)_L$
obtains a VEV. 
The fermion and scalar field content of the model and their corresponding charges under the symmetry group are listed in Table.~\ref{table:fields}. 
The left chiral fields ${Q_i}_L$ and ${L_i}_L$ transform as doublets under $SU(2)_L$ and 
are identical to the left chiral fields of the SM transforming under the $SU(2)_L$.
The right handed quarks (${u_i^0}_R, {d_i^0}_R$) form doublets under the gauge group $SU(2)_2$ as it is in case of left-right symmetric model.
But unlike the left-right symmetric model for this version of a 221 model  there are no lepton doublets under the $SU(2)_2$ gauge group and 
hence the $SU(2)_2$ is leptophobic in nature. Therefore the model contains exotic quarks to ensure triangle anomaly cancellation. 
Each generation of the exotic quark sector of the model contains a field ${XQ_i}_L$ formed out of two left chiral fields ${xu_i^0}_L$ and ${xd_i^0}_L$, and
transforms as a doublet under the $SU(2)_2$ gauge group. 
For each generation the model also contains two right chiral fields ${xu_i^0}_R$ and ${xd_i^0}_R$ which are singlets under both $SU(2)_L$ and $SU(2)_2$.
The right handed leptons ${e_i}_R$ transform as singlets under both the $SU(2)$ gauge groups.

\begin{table}[t]
\centering
  \begin{tabular}{|c|c||c|c|}
   \hline
  ${Q_i}_L$ = $\begin{pmatrix} {u_i^0}_L \\ {d_i^0}_L \end{pmatrix}$ & $ (3,2,1,\frac{1}{6})$ &   ${Q_i}_R$ = $\begin{pmatrix} {u_i^0}_R \\ {d_i^0}_R \end{pmatrix}$ & $(3,1,2,\frac{1}{6})$\\
   \hline
  {\multirow{2}{*}{${XQ_i}_L$ = $\begin{pmatrix} {xu_i^0}_L \\ {xd_i^0}_L \end{pmatrix}$}} & {\multirow{2}{*}{$(3,1,2,\frac{1}{6})$}} &${xu_i^0}_R$ & $(3,1,1,\frac{2}{3})$ \\
                                                                           &                                          &${xd_i^0}_R$ & $(3,1,1,-\frac{1}{3})$\\
   \hline
   ${L_i}_L = \begin{pmatrix} \nu_L \\ {e_i}_L \end{pmatrix}$ & $(1,2,1,-\frac{1}{2})$ &${e_i}_R$ &$(1,1,1,-1) $ \\
 \hline
 {\multirow{2}{*}{$\Phi$}} & {\multirow{2}{*}{$(1,2,2,0)$}} & $H_1$ & $(1,2,1,-\frac{1}{2})$ \\
                           &                                & $H_2$ & $(1,1,2,-\frac{1}{2})$ \\
 \hline
 \end{tabular}
 \caption{Fields and their corresponding charges under the gauge group $SU(3)_C\times SU(2)_1\times SU(2)_2\times U(1)_X$.
 The model contains three generations of fermions for $i = 1,2,3$.}
\label{table:fields}
\end{table}

Note that the exotic quarks present in the model are chiral in nature under the full unbroken gauge group. However, after the stage-I 
symmetry breaking $SU(2)_2 \times U(1)_X \rightarrow U(1)_Y$, the exotic quarks become vector-like under the $SU(2)_L \times U(1)_Y$, which is the SM gauge group.
This feature can be realized by following the definition of hypercharge quantum number $Y$ after the first stage of symmetry breaking which is given by 
\begin{equation}
 Y = T_{2_3} + Q_X,
\end{equation}
where $T_{2_3}$ denote the diagonal generator of $SU(2)_2$ and $Q_X$ represents the charge for the gauge group $U(1)_X$.
The hypercharge quantum number for the exotic quarks are given by $Y(xu_L^0) = \frac{2}{3} = Y(xu_R^0)$ 
and $Y(xd_L^0) = -\frac{1}{3} = Y(xd_R^0)$, i.e. they are vector-like with respect to the $SU(2)_L \times U(1)_Y$ gauge group.

After the stage II of symmetry breaking, the electric charge for a field is given by
\begin{align}
 Q = T_{1_3} + Y,
\end{align}
where $T_{1_3}$ denote the diagonal generator of $SU(2)_1$. Following the above definition the electric charges for the VLQs are given by 
$Q(xu_i^0) = +\frac{2}{3}$ and $Q(xd_i^0) = -\frac{1}{3}$.

\subsection{Yukawa Sector}

The Yukawa Lagrangian including the bilinear mass terms for the model is given by 
\begin{align}\label{Lag:Yukawa}
- L_{Yukawa} =\,\,& Y^q_{ij} \,\, \overline{{Q_i}}_L \,\, \Phi\,\, {Q_j}_R + Y^{qC}_{ij} \,\, \overline{{Q_i}}_L \,\, \widetilde{\Phi}  \,\, {Q_j}_R  + Y^{xqxu}_{ij} \overline{{XQ_i}}_L\,\,{xu_j^0}_R\,\,H_2 \nn \\
                  & -  Y^{xqxd}_{ij} \,\,\, \overline{{XQ_i}}_L\,\,{xd_j^0}_R\,\,\widetilde{H_2} + Y^{qxu}_{ij} \,\,\overline{{Q_i}}_L \,\,{xu_j^0}_R \,\, H_1 -  Y^{qxd}_{ij} \,\,\overline{{Q_i}}_L \,\,{xd_j^0}_R \,\, \widetilde{H_1} \nn \\ 
               & + \mu_{ij} \,\,\overline{{XQ_i}}_L\,\,{Q_j}_R  
               + Y^L_{ij}\,\, \overline{{L_i}}_L \,\, {e_j}_R\,\,\widetilde{H_1}  + H.C. ,
\end{align}
where $i,j = 1,2,3$ and we define the fields, 
\begin{align}
\widetilde{\Phi} = \sigma_2 \Phi^* \sigma_2 \equiv (1,2,2,0), \quad
\widetilde{H_1} = i \sigma_2 H_1^*\equiv (1,2,1,\frac{1}{2}), \quad
\widetilde{H_2} = i \sigma_2 H_2^*\equiv (1,1,2,\frac{1}{2}).
\end{align}
The $Y$ matrices in the above Lagrangian are Yukawa couplings. Note that after the scalars get VEVs the
$Y_{ij}^q$ and $Y_{ij}^{qC}$ terms will give masses to the SM quarks while the $Y_{ij}^{xqxu}$ and $Y_{ij}^{xqxd}$ terms give masses 
to the VLQs. 
The terms in the Lagrangian containing $Y_{ij}^{qxu}$, $Y_{ij}^{qxd}$ and the bilinear term with $\mu_{ij}$ will generate mixing between the SM quarks and the VLQs. Since the model does not contain 
any lepton doublet under $SU(2)_2$ gauge group, the charged leptons will get mass from the VEV of the 
doublet $H_1$ unlike the quarks which get their mass from the bi-doublet $\Phi$.

\subsection{Scalar Sector}
The tree level scalar potential for the model in terms of a complete set of linearly independent gauge invariant terms is given by  
\begin{align}
  V =  &-\mu_1^2 \, \textrm{Tr}[\Phi^\dag \Phi] - \Big\{\mu_2^2 \, \textrm{Tr}[\widetilde{\Phi}^\dag \Phi] + {\mu_2^2}^* \,\textrm{Tr}[\Phi^\dag \widetilde{\Phi}] \Big\}
- \mu_3^2 \, H_1^\dag H_1 - \mu_4^2 \, H_2^\dag H_2 \nonumber \\
      &+ \Big\{M_1 \, H_1^\dag \Phi H_2 + M_1^* \, H_2^\dag \Phi^\dag H_1\Big\}
       +   \Big\{M_2 \, H_1^\dag \widetilde{\Phi} H_2 + M_2^* \, H_2^\dag \widetilde{\Phi}^\dag H_1\Big\}\nonumber \\
      &+ \lambda_1\Big\{ \, \textrm{Tr}[\Phi^\dag \Phi]\Big\}^2
        + \Big\{\lambda_2 \, \{\textrm{Tr}[\Phi^\dag \widetilde{\Phi}]\}^2  + \lambda_2^* \, \{\textrm{Tr}[\widetilde{\Phi}^\dag \Phi]\}^2\Big\}
  + \lambda_3 \, \textrm{Tr}[\Phi^\dag \widetilde{\Phi}] \, \textrm{Tr}[\widetilde{\Phi}^\dag \Phi]\nonumber\\
      &+ \Big\{\lambda_4 \, \textrm{Tr}[\Phi^\dag \Phi] \, \textrm{Tr}[\Phi^\dag \widetilde{\Phi}] + \lambda_4^* \, \textrm{Tr}[\Phi^\dag \Phi] \, \textrm{Tr}[\widetilde{\Phi}^\dag \Phi]\Big\}
+ \beta_1 \, \textrm{Tr}[\Phi^\dag \Phi] (H_1^\dag H_1) \nn \\
&+ \Big\{\beta_2 \, \textrm{Tr}[\widetilde{\Phi}^\dag \Phi] (H_1^\dag H_1) + \beta_2^* \, \textrm{Tr}[\Phi^\dag \widetilde{\Phi}] (H_1^\dag H_1)\Big\} 
    + \beta_3\, \widetilde{H_1}^\dag \Phi \Phi^\dag  \widetilde{H_1} \nn \\
           &+ \alpha_1 \, \textrm{Tr}[\Phi^\dag \Phi] (H_2^\dag H_2) 
           + \Big\{\alpha_2 \, \textrm{Tr}[\widetilde{\Phi}^\dag \Phi] (H_2^\dag H_2) + \alpha_2^* \, \textrm{Tr}[\Phi^\dag \widetilde{\Phi}] (H_2^\dag H_2)\Big\}\nonumber\\
           &+ \alpha_3 \, \widetilde{H_2}^\dag \Phi^\dag \Phi \widetilde{H_2}
          + \rho_1 \, (H_1^\dag H_1)^2 + \rho_2 \, (H_2^\dag H_2)^2 + \rho_3 \, (H_1^\dag H_1)(H_2^\dag H_2).
  \end{align}
In general the parameters from the set $\{\mu_1^2, \, \mu_3^2, \, \mu_4^2, \, \lambda_1, \, \lambda_3, \, \beta_1, \, \beta_3, \, \alpha_1, \, \alpha_3, \, \rho_1, \, \rho_2, \, \rho_3\}$ 
are real and others are complex. In this paper, for simplicity we have considered all the parameters to be real. 

The scalar fields in component form  can be written as
\begin{align}\label{eqn:scalar_fields}
 &H_1 = \begin{pmatrix} \chi^0 \\ \chi^-\end{pmatrix}, \quad\quad\quad\qquad H_2 = \begin{pmatrix} \chi^{\prime 0} \\ \chi^{\prime -}\end{pmatrix}, \quad\quad\quad
 \Phi=\begin{pmatrix}\phi_1^{0} & \phi_1^{+} \\ \phi_2^{-} & \phi_2^{0} \end{pmatrix}, 
 \end{align}
where 
\begin{align}
&\phi_1^0 = \frac{1}{\sqrt{2}}(v_1 + {\phi_1^{0}}^r + i \, {\phi_1^{0}}^i ), \quad \phi_2^0 = \frac{1}{\sqrt{2}}(v_2 + {\phi_2^{0}}^r + i \, {\phi_2^{0}}^i ), \nn \\
&\chi^0 = \frac{1}{\sqrt{2}}(v_3 + {\chi^{0}}^r + i \, {\chi^{0}}^i ), \quad {\chi^\prime}^0 = \frac{1}{\sqrt{2}}(u + {{\chi^\prime}^{0}}^r + i \, {{\chi^\prime}^{0}}^i ).
\end{align}
The structure of VEVs for the scalar fields are 
\begin{align}\label{eqn:VEVs}
 &\VEV{H_1} = \mfrac{1}{\sqrt{2}} \begin{pmatrix} v_3 \\ 0 \end{pmatrix}, \quad\quad
 \VEV{H_2} = \mfrac{1}{\sqrt{2}} \begin{pmatrix} u \\ 0 \end{pmatrix}, \quad\quad
 \VEV{\Phi} = \mfrac{1}{\sqrt{2}} \begin{pmatrix} v_1 & 0 \\ 0 & v_2 \end{pmatrix}.
\end{align}

The set of tadpole equations 
$\big\{\frac{\partial V}{\partial {\phi_1^0}^r} = 0, \frac{\partial V}{\partial {\phi_2^0}^r} = 0, \frac{\partial V}{\partial {\chi^0}^r} = 0, 
\frac{\partial V}{\partial {{\chi^\prime}^0}^r} = 0\big\}$ have been solved in terms of $\mu_1^2, \, \mu_2^2, \, \mu_3^2, \, \mu_4^2$ and are given 
in the appendix \ref{appendix:tadpole}.
The components of the mass square matrices for the CP even scalars ($M_S^2$) in the $({\phi_1^0}^r,\, {\phi_2^0}^r, \, {{\chi^\prime}^0}^r, \, {{\chi^0}^r})$ basis,
for the CP odd scalars ($M_P^2$) in the $({\phi_1^0}^i,\, {\phi_2^0}^i, \, {{\chi^\prime}^0}^i, \, {{\chi^0}^i})$ basis and 
for the charged scalars ($M_C^2$) in the $(\phi_1^+, \, \phi_2^+, \, {\chi^\prime}^+, \, {\chi^+})$ basis are given in
appendix \ref{appendix:cp_even_mass_matrices}, \ref{appendix:cp_odd_mass_matrices} and \ref{appendix:charged_scalar_mass_matrices} respectively.
 
There are four physical CP even scalars in the model. Two of the four CP odd scalars will be physical and the other two are massless Goldstone bosons
which become part of the two massive neutral gauge bosons.
Similarly there will be two physical charged scalars and the other two Goldstone bosons become part of the two massive charged gauge bosons.
Hence, after the spontaneous symmetry breaking followed by the Higgs mechanism, the physical spectrum of the scalar sector consists of four neutral CP even scalars, two neutral CP odd scalars and two charged scalars (and their antiparticles).

\subsection{Gauge Boson Sector}
The gauge couplings corresponding to the gauge groups $SU(2)_L$, $SU(2)_2$ and $U(1)_X$ are respectively represented by $g_1$, $g_2$ and $g_X$.
Based on the two stages of symmetry breaking pattern we define two mixing angles $\phi$ and $\theta_W$ in terms of which the gauge couplings are given by
\begin{equation}
 g_1 = \frac{e}{\sin \theta_W}, \hspace{1cm} g_2 = \frac{e}{\cos \theta_W \sin \phi}, \hspace{1cm} g_X = \frac{e}{\cos \theta_W \cos \phi}.
\end{equation}
After the stage I of symmetry breaking $SU(2)_2\times U(1)_X\rightarrow U(1)_Y$ the gauge group becomes the Standard Model gauge group $SU(2)_L\, \times U(1)_Y$. 
The SM hypercharge gauge coupling $g_Y$ for the gauge group $U(1)_Y$ is given by the relation $\dfrac{1}{g_Y^2} =  \dfrac{1}{g_2^2} + \dfrac{1}{g_X^2}$. 
With the stage II of symmetry breaking $SU(2)_1\times U(1)_Y\rightarrow U(1)_{em}$ the electromagnetic gauge coupling constant $e$ is defined by 
$\dfrac{1}{e^2} = \dfrac{1}{g_1^2} + \dfrac{1}{g_Y^2}$. Here the angle $\theta_W$ denotes the weak mixing angle in the SM.
At the end of the two stages of symmetry breaking the electromagnetic charge for any field in the model is defined by $ Q = T_{1_3} + T_{2_3} + Q_{X}$. 

The gauge bosons corresponding to the gauge groups are denoted by :
\begin{align}
 SU(2)_1 &: W_{1,\mu}^{\pm}, W_{1,\mu}^{3}; \nonumber \\
 SU(2)_2 &: W_{2,\mu}^{\pm}, W_{2,\mu}^{3}; \nonumber \\
 U(1)_X &: X_\mu.
\end{align}
The mass square matrix for the charged gauge boson sector in the $(W_{1,\mu}^\pm, W_{2,\mu}^\pm)$ basis is given by 
\begin{equation}
 M^2_{W_1-W_2} =  \frac{1}{4}\begin{pmatrix}
                 g_1^2 \,(v_1^2 + v_2^2 + v_3^2) & -\,2\, g_1 g_2 v_1 v_2 \\
                   -\,2\, g_1 g_2 v_1 v_2 & g_2^2\, (u^2 + v_1^2 + v_2^2)\end{pmatrix}.
\end{equation}
Since the vacuum expectation value $v_3$ gives masses to the charged leptons, for simplicity
we consider the situation where $v_1^2 + v_2^2 >> v_3 ^2$ and to have the stage II breaking at a higher scale we consider 
$u >> v_1, v_2, v_3$. Based on this we define a small parameter $\epsilon$ which is given by $\dfrac{v^2}{u^2}$,
where $v = \sqrt{v_1^2 + v_2^2 + v_3^2} \simeq 246$ GeV.

The mass eigenstates for the charged gauge boson sector in terms of the gauge eigenstates are given by
\begin{equation}
 \begin{pmatrix}  W^\pm \\ {W^\prime}^\pm \end{pmatrix} 
 = 
 \begin{pmatrix}
  \cos \theta_{\text{ww}^\prime} & \sin \theta_{\text{ww}^\prime} \\
  -\sin \theta_{\text{ww}^\prime} & \cos \theta_{\text{ww}^\prime}
 \end{pmatrix}
  \begin{pmatrix}  W_1^\pm \\ W_2^\pm \end{pmatrix} ,
\end{equation}
with the mixing angle given up to order $\epsilon$ by the relation
\begin{equation}\label{eqn:wwp_mixing_angle}
  \cos \theta_{\text{ww}^\prime} \simeq 1 \hspace{1cm}  \text{and} \hspace{1cm}
  \sin \theta_{\text{ww}^\prime} \simeq \epsilon \frac{\sin \phi \, \sin 2\beta }{\tan \theta_W}, 
  \hspace{0.5cm}\text{with} \hspace{0.5cm}  \beta = \text{tan}^{-1}\left(\frac{v_1}{v_2}\right).
\end{equation}
The mass eigenstate $W$ denotes the observed SM $W$ boson while the new $W^\prime$ is heavy with mass at 
TeV scale.
The mass squared matrix for the neutral gauge boson sector in the basis $(W_1^3, W_2^3, X)$ is 
\begin{equation}
 \frac{1}{4}
 \begin{pmatrix}
 g_1^2(v_1^2 + v_2^2 +v_3^2)  & -g_1 g_2 (v_1^2 + v_2^2) & - g_1 g_X v_3^2  \\
 -g_1 g_2 (v_1^2 + v_2^2)  & g_2^2(v_1^2 + v_2^2 +u^2) & - g_2 g_X u^2 \\
 - g_1 g_X v_3^2 & - g_2 g_X u^2 & g_X^2 (u^2 + v_3^2)
 \end{pmatrix},
\end{equation}

and the mass eigenstates are given by 
\begin{align}
 \begin{pmatrix} Z^\prime \\ Z \\ A \end{pmatrix}
 = \begin{pmatrix} \cos \theta_{zz^\prime} & -\sin \theta_{zz^\prime} & 0 \\ 
                   \sin \theta_{zz^\prime} & \cos \theta_{zz^\prime} & 0 \\
                                      0 & 0 & 1
   \end{pmatrix}
   \begin{pmatrix}
       1 & 0 & 0\\
    0 & \cos\theta_W  & -\sin\theta_W \\
    0 & \sin\theta_W  & \cos\theta_W 
   \end{pmatrix}
      \begin{pmatrix}
     \cos\phi &  0 & -\sin\phi \\
           0 & 1 & 0\\
       \sin\phi & 0 &\cos\phi
      \end{pmatrix}
      \begin{pmatrix} W_2^3 \\ W_1^3 \\ X     \end{pmatrix}.
    \end{align}
 Among the mass eigenstates, $A$ denotes massless photon, $Z$ denotes the observed neutral heavy SM weak gauge boson and $Z^\prime$, the heavier neutral gauge boson that has mass at TeV scale.
Up to order $\epsilon$ the $Z-Z^\prime$ mixing angle is given by 
\begin{align}\label{eqn:zzp_mixing_angle}
\cos\theta_{zz^\prime} \simeq 1 \,\,\,\,\,\text{and}\,\,\,
 \sin\theta_{zz^\prime} \simeq \epsilon \,\frac{\sin\phi \, \cos^3\phi}{\sin\theta_W}.
\end{align}

\section{Interactions of Vector Like Quarks with Gauge Bosons}\label{section:Int_With_Gauge_Bosons}
The covariant derivative for a field determines its nature of interaction with the gauge bosons. To observe the vector-like nature of the interaction of exotic quarks with the 
SM gauge bosons we write those terms from the covariant derivative which contains neutral gauge bosons and that is given by
\begin{align}\label{eqn:neutral_boson_interaction}
&g_1\,T_{1_3}\,{W_1^3}_\mu +  g_2\,T_{2_3}\,{W_2^3}_\mu + g_X \ Q_x \, X_\mu \nn \\
=\,\,& e \, Q \, A_\mu \, \, + \,\,\,\frac{e}{\sin\theta_W\,\cos\theta_W}  \Big\{ (T_{1_3} - Q\, \sin^2\theta_W) + \epsilon \, \cos^2\phi\, \,\big((T_{1_3}-Q)\sin^2\phi + T_{2_3}\big) \Big\} \, Z_\mu \nn \\
\,\,\,\,+&\,\,\,\frac{e}{\sin\theta_W\,\cos\theta_W}  \Big\{ \big((T_{1_3}-Q)\sin^2\phi + T_{2_3}\big)\frac{\sin\theta_W}{\sin\phi \, \cos\phi} + \epsilon \, (Q\,\sin^2\theta_W - T_{1_3})\,\frac{\sin\phi \cos^3\phi}{\sin\theta_W} \Big\} \, Z^\prime_\mu.
\end{align}
By observing the term with $Z_\mu$ from Eq.~\ref{eqn:neutral_boson_interaction} we can conclude that the interaction strength of 
the left chiral fields ${xu_i^0}_L$(${xd_i^0}_L$) with $Z$ boson differs from the interaction strength for the right chiral fields ${xu_i^0}_R$(${xd_i^0}_R$) by a  term proportional to
$\epsilon$. This is because the left chiral and the right chiral fields do not have same $T_{2_3}$ value.
 Hence the interactions of the BSM quarks with the $Z$ boson are vector-like in the limit $\epsilon\rightarrow 0$. 
 The origin of the $\epsilon$ term is due to the $Z-Z^\prime$ mixing. 
 Since we are interested in the situation where $u\gg v$ and hence $\epsilon = \frac{v^2}{u^2} \ll 1$,
 we have identified the BSM exotic quarks ($xu_i^0$ and $xd_i^0$) as VLQs throughout the article.
 
 Since the exotic quarks are singlets under $SU(2)_L$ the interaction strength of the exotic quarks with the $W$ boson will be very small for small values of $W-W^\prime$ mixing angle.

\section{FCNC in the presence of Vector Like Quarks}\label{section:FCNC}
In SM there is no flavour-changing neutral current (FCNC) interactions of quarks with the $Z$ boson at the tree level
because 
the quarks with same electric charge have universal charge assignments under the SM gauge group. But in models with VLQs this scenario 
of having universal charges under the gauge group of the model having the same electric charge breaks down. Hence the mixing between the SM quarks 
and the VLQs can generate tree level FCNC interactions for the SM quarks.

In our model this mixing is generated by the terms proportional to $Y^{qxd}_{ij}$, $Y^{qxu}_{ij}$ and $\mu_{ij}$ in the Yukawa Lagrangian 
in Eq.~\ref{Lag:Yukawa}.
For models with vector like quarks the possibility of restricting tree level FCNC interactions exists for special choice of mixing patterns 
between the quarks and VLQs\cite{Langacker:1988ur}. 
It has been shown in \cite{Langacker:1988ur} that if one linear combination of VLQs mix
with only one SM quark mass eigenstate then there will be no $Z$-boson mediated FCNC interaction at tree level. This would imply that 
each VLQ will have a corresponding SM quark (mass eigenstate) with which it mixes.
We use this formalism \cite{Langacker:1988ur} and discuss the scenario in which the FCNC interactions vanish for our model. 

The $6\times6$ dimensional mass matrices for the up-quark sector and for the down-quark sector are respectively given by
\begin{equation}\label{Matrix:up_down_sector}
 \mathcal{M}^u = \begin{pmatrix}
  M^u & Y^{qxu} \frac{v_3}{\sqrt{2}}\\
  \mu & Y^{xqxu} \frac{u}{\sqrt{2}}
 \end{pmatrix} 
 \hspace{1cm} \text{and} \hspace{1cm}
 \mathcal{M}^d = \begin{pmatrix}
  M^d & Y^{qxd} \frac{v_3}{\sqrt{2}}\\
  \mu & Y^{xqxd} \frac{u}{\sqrt{2}}
 \end{pmatrix}.
\end{equation}

The matrices $Y^{qxu}, \, Y^{qxd}, \, Y^{xqxd}, \mu$ are $3\times 3$ dimensional whose components are formed out of the Yukawa couplings in Eq.~\ref{Lag:Yukawa}.
The $3\times 3$ matrices $M^u$ and $M^d$ are given by 
\begin{align} 
 {M^u}_{ij} &= \frac{1}{\sqrt{2}} \big(Y^q_{ij} v_1 + Y^{qC}_{ij} v_2\big) \nonumber \\
{M^d}_{ij} &=\frac{1}{\sqrt{2}} \big( Y^q_{ij} v_2 + Y^{qC}_{ij} v_1\big).
\end{align}
The quark gauge eigenstates ($\widehat{\mathcal{U}}_{L/R}$, $\widehat{\mathcal{D}}_{L/R}$)  and the mass eigenstates 
($\mathcal{U}_{L/R}$, $\mathcal{D}_{L/R}$) are represented by  
\begin{align}\label{eqn:ucap61}
 \widehat{\mathcal{U}}_{L/R} &= \begin{pmatrix}U^0 & XU^0\end{pmatrix}^T_{L/R} \equiv\begin{pmatrix} u^0 & c^0 & t^0 & xu_1^0 & xu_2^0 & xu_3^0\end{pmatrix}^T_{L/R} \nn \\
  \widehat{\mathcal{D}}_{L/R} &= \begin{pmatrix}D^0 & XD^0\end{pmatrix}^T_{L/R}  \equiv\begin{pmatrix}d^0 & s^0 & b^0 & xd_1^0 & xd_2^0 & xd_3^0\end{pmatrix}^T_{L/R} \nn \\
   \mathcal{U}_{L/R} &= \begin{pmatrix}U & XU\end{pmatrix}^T_{L/R} \equiv\begin{pmatrix}u & c & t & xu_1 & xu_2 & xu_3\end{pmatrix}^T_{L/R} \nn \\
  \mathcal{D}_{L/R} &= \begin{pmatrix}D & XD\end{pmatrix}^T_{L/R} \equiv\begin{pmatrix}d & s & b & xd_1 & xd_2 & xd_3\end{pmatrix}^T_{L/R}\,\,.
\end{align}
The matrices $\mathcal{M}^u$ and $\mathcal{M}^d$ from Eq.~\ref{Matrix:up_down_sector}  will be diagonalized by biunitary transformations
and are given by
\begin{align}\label{up_quark_mixing_1}
 \widehat{\mathcal{U}}_{L/R} &= S_{L/R}^u \,\, \mathcal{U}_{L/R} \nn \\
 \widehat{\mathcal{D}}_{L/R} &= S_{L/R}^d \,\, \mathcal{D}_{L/R},
\end{align}
where the $S_{L}^u$ and $S_{L}^d$ are $6\times6$ unitary matrices and can be represented by
\begin{align}\label{up_quark_mixing_2}
 \mathcal{S}_L^u = \begin{pmatrix} A_L^u & E_L^u \\ F_L^u & G_L^u \end{pmatrix}, \,\,\,
 \mathcal{S}_L^d = \begin{pmatrix} A_L^d & E_L^d \\ F_L^d & G_L^d \end{pmatrix}. 
\end{align}
The matrices $\mathcal{S}_R^u$ and $\mathcal{S}_R^d$ can be obtained from Eq.~\ref{up_quark_mixing_2} by replacing 
$L \rightarrow R$.
The matrices $A, E, F, G$ are $3\times3$ dimensional and where $E$ and $F$ connect the SM quarks with the VLQs.
To avoid FCNC at the tree level we choose the mixing pattern such that the matrices for the left chiral sector take the form 
\begin{align}\label{up_quark_mixing_3}
 A_L^u = \widehat{A_L^u} C_L^u, \, F_L^u = S_L^u, \, G_L^u = C_L^u, \, E_L^u = -\widehat{A_L^u} S_L^u,
\end{align}
where 
\begin{align}\label{up_quark_mixing_4}
&{\widehat{A_L^u}}^\dagger \widehat{A_L^u} = \widehat{A_L^u}  {\widehat{A_L^u}}^\dagger = \mathbb{1},\nn \\ 
&C_L^u = \text{diag}(\cos \theta_L^u, \, \cos\theta_L^c, \cos\theta_L^t), \,
S_L^u = \text{diag}(\sin \theta_L^u, \, \sin\theta_L^c, \sin\theta_L^t),
\end{align}
then, 
\begin{align}\label{up_quark_mixing_5}
  u_L &= \cos \theta_L^u  \Bigg(\widehat{A_L^u}^\dagger \begin{pmatrix} u^0 \\ c^0 \\ t^0 \end{pmatrix}_L\Bigg)_1 + \sin \theta_L^u xu_{1L}^0, \nn \hspace{0.5cm}
 {xu_1}_L = - \sin \theta_L^u \Bigg(\widehat{A_L^u}^\dagger \begin{pmatrix} u^0 \\ c^0 \\ t^0 \end{pmatrix}_L\Bigg)_1 +  \cos \theta_L^u \, \, \, xu_{1L}^0, \nn \\
  c_L &= \cos \theta_L^c  \Bigg(\widehat{A_L^u}^\dagger \begin{pmatrix} u^0 \\ c^0 \\ t^0 \end{pmatrix}_L\Bigg)_2 + \sin \theta_L^c \, \, \, xu_{2L}^0, \nn \hspace{0.5cm}
 {xu_2}_L = - \sin \theta_L^c \Bigg(\widehat{A_L^u}^\dagger \begin{pmatrix} u^0 \\ c^0 \\ t^0 \end{pmatrix}_L\Bigg)_2 + \cos \theta_L^c \, \, \, xu_{2L}^0, \nn \\
   t_L &= \cos \theta_L^t \Bigg(\widehat{A_L^u}^\dagger \begin{pmatrix} u^0 \\ c^0 \\ t^0 \end{pmatrix}_L\Bigg)_3 + \sin \theta_L^t \, \, \, xu_{3L}^0, \nn \hspace{0.5cm}
 {xu_3}_L = - \sin \theta_L^t \Bigg(\widehat{A_L^u}^\dagger \begin{pmatrix} u^0 \\ c^0 \\ t^0 \end{pmatrix}_L\Bigg)_3 +  \cos \theta_L^t \, \, \, xu_{3L}^0. \\
\end{align}
From Eq.~\ref{up_quark_mixing_5} it can be seen that each vector like quark mixes with only one linear combination of the SM gauge eigenstate quarks. 
The different linear combinations with which different VLQs mix are characterized by the unitary matrix $\widehat{A_L^u}$.  

Similarly to avoid FCNC in the down-type quark sector we choose the  mixing matrices for the left chiral down-type quarks in a similar way as above with up-types changed with down-type quarks : 
\begin{align}\label{down_quark_mixing_1}
 A_L^d = \widehat{A_L^d} C_L^d, \hspace{0.5cm} F_L^d = S_L^d, \,\hspace{0.5cm} G_L^d = C_L^d, \hspace{0.5cm} E_L^d = -\widehat{A_L^d} S_L^d
\end{align}
where 
\begin{align}\label{down_quark_mixing_2}
&{\widehat{A_L^d}}^\dagger \widehat{A_L^d} = \widehat{A_L^d}  {\widehat{A_L^d}}^\dagger = \mathbb{1},\nn \\ 
&C_L^d = \text{diag}(\cos \theta_L^d, \, \cos\theta_L^s, \cos\theta_L^b), \hspace{0.5cm} 
S_L^d = \text{diag}(\sin \theta_L^d, \, \sin\theta_L^s, \sin\theta_L^b)\,\,.
\end{align}
The mixing matrices and the mass eigenstates for the right handed fields can be obtained by replacing $L\rightarrow R$ in
Eqs. \ref{up_quark_mixing_3}-\ref{down_quark_mixing_2}. Note that  both Eq.~\ref{up_quark_mixing_5} and its right handed counterpart
show that in the absence of mixing between the SM quarks and the VLQs, the unitary matrices $\widehat{A_L^u}$ and $\widehat{A_R^u}$ 
are the matrices which diagonalize the mass matrix for the up-quark sector of the SM. The same can be concluded for the 
down-quark sector also.
\subsection{CKM Matrix in presence of Vector Like Quarks}
The interaction of the SM left-chiral gauge eigenstates with the $W$ boson is given by 
\begin{align}
\frac{g_1}{\sqrt{2}}W_\mu^+ \, \overline{U_L^0} \, \gamma^\mu \, D_L^0 = \frac{g_1}{\sqrt{2}} W_\mu^+ \Big\{ &\overline{U_L} \, {A_L^u}^\dagger A_L^d \gamma^\mu \, D_L  
                                                                                             +  \overline{XU_L} \, {E_L^u}^\dagger A_L^d \gamma^\mu D_L \nn \\
                                                                                             & + \overline{U_L} \, {A_L^u}^\dagger E_L^d \gamma^\mu \, XD_L
                                                                                             + \overline{XU_L} \, {E_L^u}^\dagger E_L^d \gamma^\mu \, XD_L\Big\}.
\end{align}
Based on the interactions of the SM quark mass eigenstates with the $W$ boson the CKM matrix is defined as 
\begin{equation}
 V_L^{CKM} = {A_L^u}^\dagger A_L^d = C_L^u \, \widehat{{A_L^u}}^\dagger \, \widehat{A_L^d} \, C_L^d.
\end{equation}
$V_L^{CKM}$ corresponds to the measured CKM matrix. It can be noted that in the presence of mixing between SM quarks and VLQs the matrix $V_L^{CKM}$ is not unitary.
The deviation from unitarity of the measured CKM matrix will put constraints on the mixing angles contained 
in the matrices $C_L^u$ and $C_L^d$. 

Similarly the interaction term of the SM right-chiral quark gauge eigenstates with the $W^\prime$ gauge boson is given by
 \begin{align}
\frac{g_2}{\sqrt{2}}{W^\prime}_\mu^+ \, \overline{U_R^0} \, \gamma^\mu \, D_R^0 = \frac{g_1}{\sqrt{2}} {W^\prime}_\mu^+ \Big\{ &\overline{U_R} \, {A_R^u}^\dagger A_R^d \gamma^\mu \, D_R  
                                                                                             +  \overline{XU_R} \, {E_R^u}^\dagger A_R^d \gamma^\mu D_R \nn \\
                                                                                             & + \overline{U_R} \, {A_R^u}^\dagger E_R^d \gamma^\mu \, XD_R
                                                                                             + \overline{XU_R} \, {E_R^u}^\dagger E_R^d \gamma^\mu \, XD_R\Big\}.
\end{align}
We define a right-handed CKM matrix given by 
\begin{equation}
V_R^{CKM} = {A_R^u}^\dagger A_R^d = C_R^u \, \widehat{{A_R^u}}^\dagger \, \widehat{A_R^d} \, C_R^d.
\end{equation}

\subsection{FCNC interaction with $Z$ and $Z^\prime$}
To see how the choice of mixing pattern that we have considered avoids FCNC at tree level, we focus on the terms containing $Z$ boson in Eq.~\ref{eqn:neutral_boson_interaction}. 
Since all the fields $u^0_L$, $c^0_L$ and $t^0_L$ carry universal charges under the full gauge group, we can write their interaction with the $Z$ boson in terms of 
$U^0$ of Eq.~\ref{eqn:ucap61} as
\begin{equation}
\frac{e}{\sin\theta_W\,\cos\theta_W}  \Big\{ \big(\frac{1}{2} - \frac{2}{3}\, \sin^2\theta_W\big) - \frac{1}{6}\, \epsilon \, \cos^2\phi\, \,\sin^2\phi  \Big\} \, Z_\mu \, \overline{U_L^0} \, \gamma^\mu \, U_L^0.
\end{equation}

The flavour diagonal nature can be seen by writing $\overline{U_L^0} \, \gamma^\mu \, U_L^0$ in terms of mass eigenstates by using
the Eqs. \ref{eqn:ucap61}-\ref{up_quark_mixing_3} and is given by
\begin{equation}\label{eqn:fcnc2}
\overline{U_L^0} \, \gamma^\mu \, U_L^0 = \, \overline{U_L} \, {C_L^u}^2\, \gamma^\mu \, U_L 
- \,(\overline{U_L} \, C_L^u\,S_L^u \,\gamma^\mu \, XU_L\, + \overline{XU_L} \, S_L^u\,C_L^u \, \gamma^\mu \, U_L \,) 
+  \, \overline{XU_L} \, {S_L^u}^2\,\gamma^\mu \, XU_L.
\end{equation}

Since $C_L^u$ is a diagonal matrix, the first term in the right hand side of the above equation is diagonal in the mass eigenstates $u_L$, $c_L$ and $t_L$. 
Similarly, with the equivalent form of Eq.\ref{eqn:fcnc2} for $U_R^0$, $XU_{L/R}^0$, $D_{L/R}^0$ and $XD_{L/R}^0$, and since $S_L^u$ is 
also a diagonal matrix, we find that there is no flavour-changing interactions of the SM quarks with the $Z$ boson. 
Again, by expanding the interaction terms for $Z^\prime$ in Eq.~\ref{eqn:neutral_boson_interaction} we find that
the $Z^\prime$ also does not have any flavour-changing interactions with the SM quarks for the chosen mixing pattern.

\subsection{FCNH interaction with Higgs Bosons}
The unitary matrices in Eq.~\ref{up_quark_mixing_1} diagonalize the quark mass matrices, i.e.,
\begin{align}\label{quark_matrix_diagonalization}
 {\mathcal{S}_L^u}^\dagger \, \begin{pmatrix} \frac{1}{\sqrt{2}}(Y^q \, v_1 + Y^{qC} \,v_2^*)  & Y^{qxu} \frac{v_3}{\sqrt{2}} \\
                                    \mu & Y^{xqxu} \frac{u}{\sqrt{2}}
                    \end{pmatrix} \mathcal{S}_R^u = \mathcal{M}^u_{diag}\,\,\, ,                    \nn  \\
 {\mathcal{S}_L^d}^\dagger \, \begin{pmatrix} \frac{1}{\sqrt{2}}(Y^q \, v_2 + Y^{qC}\, v_1^*)  & Y^{qxd} \frac{v_3}{\sqrt{2}} \\
                                    \mu & Y^{xqxd} \frac{u}{\sqrt{2}}
                    \end{pmatrix} \mathcal{S}_R^d = \mathcal{M}^d_{diag}\,\, ,                
\end{align}
where
\begin{align}
 \mathcal{M}^u_{diag} = \begin{pmatrix} M_{diag}^u & 0 \\ 0 & M_{diag}^{xu} \end{pmatrix} \quad \text{and} \quad
  \mathcal{M}^d_{diag} = \begin{pmatrix} M_{diag}^d & 0 \\ 0 & M_{diag}^{xd} \end{pmatrix}.
\end{align}
And 
\begin{align}
  M_{diag}^u = \text{diag}(m_u,\, m_c,\, m_t), \hspace{0.5cm} M_{diag}^d = \text{diag}(m_d,\, m_s,\, m_b),\hspace{2cm}\nn \\
M_{diag}^{xu} = \text{diag}(m_{xu_1},\, m_{xu_2},\, m_{xu_3}), \hspace{0.5cm} M_{diag}^{xd} = \text{diag}(m_{xd_1},\, m_{xd_2},\, m_{xd_3}).
\end{align}

From Eq.~\ref{quark_matrix_diagonalization} the Yukawa couplings with the bi-doublet in terms of mixing matrices are given by
\begin{align}\label{soln_yukawa_coupling_1}
 \frac{1}{\sqrt{2}} (Y^q \, v_1 + Y^{qC} \,v_2^*) = A_L^u\, M_{diag}^u \,{A_R^u}^\dagger + E_L^u \, M_{diag}^{xu} \,{E_R^u}^\dagger, \nn \\
 \frac{1}{\sqrt{2}} (Y^q \, v_2 + Y^{qC} \,v_1^*) = A_L^d\, M_{diag}^d \,{A_R^d}^\dagger + E_L^d \, M_{diag}^{xd} \,{E_R^d}^\dagger. 
\end{align}

Solving Eq.~\ref{soln_yukawa_coupling_1} for $Y^q$ and $Y^{qC}$ we get
\begin{align}\label{soln_yukawa_coupling_2}
 Y^q &= \frac{\sqrt{2}}{|v_1|^2 -|v_2|^2} \bigg(v_1^* (A_L^u\, M_{diag}^u \,{A_R^u}^\dagger + E_L^u \, M_{diag}^{xu} \,{E_R^u}^\dagger) 
                                          - v_2^* (A_L^d\, M_{diag}^d \,{A_R^d}^\dagger + E_L^d \, M_{diag}^{xd} \,{E_R^d}^\dagger)\bigg), \nn \\
 Y^{qC} &= \frac{\sqrt{2}}{|v_1|^2 -|v_2|^2} \bigg(- v_2 (A_L^u\, M_{diag}^u \,{A_R^u}^\dagger + E_L^u \, M_{diag}^{xu} \,{E_R^u}^\dagger) 
                                          + v_1 (A_L^d\, M_{diag}^d \,{A_R^d}^\dagger + E_L^d \, M_{diag}^{xd} \,{E_R^d}^\dagger)\bigg).
\end{align}
From the Yukawa Lagrangian in Eq.~\ref{Lag:Yukawa} the interaction of the neutral components of the bi-doublet $\Phi$ with the SM quarks in the gauge basis is given by 
\begin{equation}
   \overline{U_L^0}\, ( Y^q \, \phi_1^0 + Y^{qC} {\phi_2^0}^*) U_R^0 \, + \, \overline{D_L^0}\, ( Y^q \, \phi_2^0 + Y^{qC} {\phi_1^0}^*) D_R^0.
\end{equation}

Using Eq.~\ref{soln_yukawa_coupling_2} and Eqs. \ref{up_quark_mixing_1}-\ref{down_quark_mixing_2} the interactions
of the scalars $\phi_1^0$ and $\phi_2^0$ with the SM up type quarks in the mass basis can be written as 
\begin{align}\label{eqn:up_quark_higgs_int}
 &\overline{U_L}\, {A_L^u}^\dagger \,(Y^q \, \phi_1^0 + Y^{qC} \, {\phi_2^0}^*)\, A_R^u\, U_R \nn \\
 &\,\,\,\,\,= \frac{\sqrt{2}}{v_{-}^2} \,\, \overline{U_L}\, \bigg(\, (v_1^* \phi_1^0 - v_2 {\phi_2^0}^*) \big( {C_L^u}^2 M^u_{\text{diag}} {C_R^u}^2 
                                                                              + C_L^u S_L^u M_{\text{diag}}^{xu} S_R^u C_R^u \big)\nn \\
                                  & \quad \quad\quad \, \, \, \, \, \, \, + (- v_2^* \phi_1^0 + v_1 {\phi_2^0}^*) \big( V_L^{\text{CKM}} M^d_{\text{diag}} {V_R^{CKM}}^\dagger + {A_L^u}^\dagger E_L^d M^{xd}_{\text{diag}} {E_R^d}^\dagger A_R^u \, \big) \bigg) \, \, U_R    \nn \\                                               
 &\,\,\,\,\,= \frac{\sqrt{2}}{v_{-}^2} \,\, \overline{U_L}  \, \bigg(\, \phi_{-}^0 \frac{v_{-}^2}{v_+} \big( {C_L^u}^2 M^u_{\text{diag}} {C_R^u}^2 
                                                                              + C_L^u S_L^u M_{\text{diag}}^{xu} S_R^u C_R^u \big) \nn \\
\begin{split}
                                  & \qquad\qquad\quad + \phi_+^0 \Big\{ \frac{-2 v_1^* v_2}{v_+} \big( {C_L^u}^2 M^u_{\text{diag}} {C_R^u}^2 + C_L^u S_L^u M_{\text{diag}}^{xu} S_R^u C_R^u \big) \\
                          &\qquad\qquad\,\,\,\,\,\, +  v_+ \big( V_L^{\text{CKM}} M^d_{\text{diag}} {V_R^{CKM}}^\dagger + V_L^{CKM} \inv{C_L^d} S_L^d M^{xd}_{\text{diag}} S_R^d \inv{C_R^d} {V_R^{CKM}}^\dagger \big) \Big\} \bigg) \, \, U_R , 
\end{split}
\end{align}

where the two orthogonal fields $\phi_-^0$ and $\phi_+^0$ are given by\cite{Deshpande:1990ip}
\begin{align}
\phi_+^0 &= \frac{1}{v_+} (- v_2^* \, \phi_1^0 + v_1 \,  {\phi_2^0}^*), \nn \\
\phi_{-}^0 &= \frac{1}{v_+} (v_1^* \,  \phi_1^0 + v_2 \,  {\phi_2^0}^*).
\end{align}

And $v_{\pm}^2 = |v_1|^2 \pm |v_2|^2$.
Similarly the interaction terms for the SM down type quarks in the mass basis is given by
\begin{align}\label{eqn:down_quark_higgs_int}
 &\overline{D_L}\, {A_L^d}^\dagger \,(Y^q \, \phi_2^0 + Y^{qC} \, {\phi_1^0}^*)\, A_R^d\, D_R \nn \\
 &\quad=\frac{\sqrt{2}}{v_{-}^2} \, \overline{D_L}\bigg(\, {\phi_{-}^0}^* \, \, \frac{v_{-}^2}{v_+} \big( {C_L^d}^2 M^d_{\text{diag}} {C_R^d}^2 + C_L^d S_L^d M_{\text{diag}}^{xd} S_R^d C_R^d \big) \nn \\
 \begin{split}
  & \qquad\qquad\quad\quad+ {\phi_+^0}^* \Big\{ \frac{-2 v_1 v_2^*}{v_+} \big( {C_L^d}^2 M^d_{\text{diag}} {C_R^d}^2 
                                                                              + C_L^d S_L^d M_{\text{diag}}^{xd} S_R^d C_R^d \big)  \\
  &\qquad\qquad\quad\quad + v_+ \big( {V_L^{\text{CKM}}}^\dagger M^u_{\text{diag}} {V_R^{CKM}} + {V_L^{CKM}}^\dagger \inv{C_L^u} S_L^u M^{xu}_{\text{diag}} S_R^u \inv{C_R^u} V_R^{CKM} \big) \Big\} \bigg) \, \, D_R .  
\end{split}                          
\end{align}
From Eq.~\ref{eqn:up_quark_higgs_int} and Eq.~\ref{eqn:down_quark_higgs_int} it can be concluded that the interactions
$\phi_-^0$ is flavour-diagonal but the interactions of $\phi_+^0$ is flavour-changing. $\phi_+^0$ interactions are flavour-changing because the matrix
$V_L^{CKM}$ which is the measured CKM matrix is not diagonal. Although the matrix $V_R^{CKM}$ can be non-diagonal there is no experimental 
constraint which forces it to be non-diagonal and hence $V_R^{CKM}$ can be taken to be diagonal by proper choice of Yukawa couplings. 
In left-right symmetric model the field $\phi_-^0$ is always flavour-conserving in nature \cite{Deshpande:1990ip}.
But in the 221 model we are discussing $\phi_-^0$ can also have flavour-violating interactions for general mixing patterns between VLQs 
and the SM quarks. It is the form of the mixing matrices in Eq.~\ref{up_quark_mixing_3} and Eq.~\ref{down_quark_mixing_1} which ensures that 
$\phi_-^0$ have flavour-conserving interactions.

In general $\phi_+^0$ is not a mass eigenstate and when both $v_1$ and $v_2$ are nonzero all the neutral mass eigenstates will contain $\phi_+^0$. 
And hence all of the neutral scalars have to be heavy to avoid constraints from FCNC interactions. Therefore when both $v_1$ and $v_2$ take nonzero values it will be
impossible to get a light mass eigenstate at 125 GeV which do not have flavour violating interactions. 
This scenario has been extensively discussed in the context of left-right symmetric model in \cite{Gunion:1989in}.

Hence to have a flavour-conserving Higgs at 125 GeV we made the choice $v_2 =0$, which gives
\begin{align}
v_+ = v_- = v_1, \qquad
\phi_+^0 = {\phi_2^0}^*, \qquad
\phi_-^0 = {\phi_1^0}.
\end{align}
The VEV $v_1$ has been considered as a real parameter. As we will show it is possible to choose the parameters of the scalar potential in the model such that $\phi_2^0$ do not mix with any other 
scalars and can be made heavy to avoid large neutral flavour-changing interactions. $\phi_1^0$ will be a part of the observed 125 GeV Higgs .
For $v_2 = 0$ the interactions of flavour conserving Higgs $\phi_1^0$ with the SM type quarks are given by
\begin{align}\label{eqn:quark_fc_cons_int}
  \frac{\sqrt{2}}{v_{1}} \, \overline{U_L}\bigg(&\, \phi_{1}^0 \, \big( {C_L^u}^2 M^u_{\text{diag}} {C_R^u}^2  + C_L^u S_L^u M_{\text{diag}}^{xu} S_R^u C_R^u \big)\bigg) \, \, U_R + \text{H.C.},\nn \\
  \frac{\sqrt{2}}{v_{1}} \, \overline{D_L}\bigg( &\, {\phi_{1}^0}^* \, \, \big( {C_L^d}^2 M^d_{\text{diag}} {C_R^d}^2 + C_L^d S_L^d M_{\text{diag}}^{xd} S_R^d C_R^d \big)\bigg) \, \, D_R +\text{H.C.} \, . 
\end{align}
And the interactions for the flavour-violating Higgs $\phi_2^0$ are given by
\begin{align}\label{eqn:quark_fc_viol_int}
 \frac{\sqrt{2}}{v_{1}} \, \overline{U_L}\bigg(&\,{\phi_2^0}^* \Big\{ \big( V_L^{\text{CKM}} M^d_{\text{diag}} {V_R^{CKM}}^\dagger + {A_L^u}^\dagger E_L^d M^{xd}_{\text{diag}} {E_R^d}^\dagger A_R^u \, \big) \Big\} \bigg) \, \, U_R  + H.C., \nn \\
\frac{\sqrt{2}}{v_{1}} \, \overline{D_L}\bigg( &\,{\phi_2^0} \Big\{\big( {V_L^{\text{CKM}}}^\dagger M^u_{\text{diag}} {V_R^{CKM}} + {A_L^d}^\dagger E_L^u M^{xu}_{\text{diag}} {E_R^u}^\dagger A_R^d \, \big) \Big\} \bigg) \, \, D_R   + H.C.
\end{align}

To study the nature of interactions of the neutral components from the doublets $H_1$ and $H_2$ we list the Yukawa couplings in terms of the 
mixing matrices as below (for mixing matrices of the form taken in Eq.~\ref{up_quark_mixing_3} and $v_2 = 0$)
\begin{align}\label{eqn:yukawa_up_2}
 Y^q \frac{v_1}{\sqrt{2}} &= \widehat{A_L^u} \Big(C_L^u M^u_{diag} C_R^u + S_L^u M^{xu}_{diag} S_R^u \Big) \widehat{A_R^u}^\dagger, \nn \\
 Y^{qxu}\frac{v_3}{\sqrt{2}} &= \widehat{A_L^u} \Big(C_L^u M^u_{diag} S_R^u - S_L^u M^{xu}_{diag} C_R^u\Big), \nn \\
 \mu &= (S_L^u M^u_{diag} C_R^u -C_L^u M^{xu}_{diag} S_R^u) \, \widehat{A_R^u}^\dagger, \nn \\
  Y^{xqxu}\frac{u}{\sqrt{2}} &= S_L^u M^u_{diag} S_R^u + C_L^u M^{xu}_{diag} C_R^u.
\end{align}
Similar relations for the matrices $Y^{qC}$, $Y^{qxd}$, $\mu$ and $Y^{xqxd}$ can be found by using the  mixing matrices for the down-type quark sector. Note that the mixing angles
and the mass eigenvalues should be such that both the up-quark sector and the down-quark sector yield the same $\mu$ matrix.

The fields $\chi^0$ and ${\chi^\prime}^0$ in Eq.~\ref{eqn:scalar_fields} are the neutral components of the doublets $H_1$ and $H_2$ respectively.
The interactions of the up-type SM mass eigenstate quarks with $\chi^0$ and ${\chi^\prime}^0$ can be derived from the terms proportional 
to $Y^{qxu}_{ij}$ and $Y^{xqxu}_{ij}$ respectively in the Yukawa Lagrangian. These interactions are given by
\begin{align}\label{eqn:chi_interaction}
 Y_{ij}^{qxu} \, \overline{{Q_i}_L} \, {xu_j^0}_R \, H_1  &\supset \chi^0 \,\, \overline{U_L^0} \,\, Y^{qxu} \,\, XU_R^0 \nn \\
                                                        &\supset  \frac{\sqrt{2}}{v_3} \,\, \chi^0 \,\overline{U_L} \Big[\,C_L^u \, \Big( C_L^u M^u_{diag} S_R^u - S_L^u M^{xu}_{diag} C_R^u\Big)\, S_R^u \Big]\,U_R \nn \\
 \text{and}\qquad Y_{ij}^{xqxu} \, \overline{{XQ_i}_L} \, {xu_j^0}_R \, H_2   &\supset {\chi^\prime}^0 \,\, \overline{XU_L^0}\,\, Y^{xqxu} \,\, XU_R^0 \nn \\
                                                        &\supset \frac{\sqrt{2}}{u} \,\,{\chi^\prime}^0 \,\overline{U_L} \Big[\,S_L^u \, \Big( S_L^u M^u_{diag} S_R^u + C_L^u M^{xu}_{diag} C_R^u\Big)\, S_R^u \Big]\,U_R.                                               
\end{align}
The superset sign ($\supset$) has been used in the above equations to highlight the terms containing only the fields
$\chi^0$, ${\chi^\prime}^0$ and the SM up-type mass eigenstate quarks ($U_L\, , U_R$). From Eq.~\ref{eqn:chi_interaction} it can be concluded that the scalars 
$\chi^0$ and ${\chi^\prime}^0$ do not have flavour changing interactions with the SM mass eigenstate quarks. 

Based on the above discussions, we denote the field $\phi_2^0$ as the FCNH (flavour-changing neutral Higgs) scalar and the fields
$\phi_1^0$, $\chi^0$ and ${\chi^\prime}^0$ as the three non-FCNH scalars in the model. One linear combination of the three non-FCNH 
scalars will be the 125 GeV SM Higgs and the other linear combinations can lie at the sub-TeV scale. 

The special cases for the matrices $Y^{qxu}$ and $\mu$ which will play an important role in the phenomenology of the VLQs are given by :
\begin{equation}\label{eqn:yukawa_zero}
 Y^{qxu} = 0 \implies  \tan \theta_L^i = \frac{m_i}{m_{xu_i}} \tan\theta_R^i
\end{equation}
and 
\begin{equation}
 \mu = 0 \implies \tan \theta_R^i = \frac{m_i}{m_{xu_i}} \tan\theta_L^i,
\end{equation}
where $i\in(u,c,t)$.

For simplicity we consider the scenario where the three matrices $\widehat{A_L^d}$, $\widehat{A_R^u}$ and $\widehat{A_R^d}$ are equal to $\mathbb{1}$.
The choice for the Yukawa couplings which will lead to such a scenario can be made by using $\widehat{A_L^d} = \mathbb{1}$, $\widehat{A_R^u} = \mathbb{1}$ and $\widehat{A_R^d} = \mathbb{1}$ in Eq.~\ref{eqn:yukawa_up_2}.
From Eq.~\ref{eqn:yukawa_up_2} it can be observed that the complete determination of Yukawa couplings will depend on the SM quark masses, desired values of VLQ masses, desired values of mixing angles, VEVs and on the form of the matrix $\widehat{A_L^u}$.
The measured CKM matrix will enter through the matrix $\widehat{A_L^u}$ because, $V_L^{\text{CKM}} = {A_L^u}^\dagger A_L^d = C_L^u \widehat{A_L^u}^\dagger \widehat{A_L^d} C_L^d$.

\section{Phenomenological aspects}\label{section:phenomenology}
So far we have discussed about the methodology to get rid of any tree level $Z$ boson FCNC and to achieve a 125 GeV Higgs with no flavor changing interactions. Now we shall discuss some of the phenomenological implications of the model. To do that we 
will choose a representative mass spectrum for the exotic particles in our model and then discuss the possible collider signals 
which can be explored at the LHC. 

\subsection{Scalar and gauge boson mass spectrum}
To relate the gauge eigenstates with the mass eigenstates for the scalar sector, we introduce three $4\times 4$ matrices $Z^E$, $Z^O$ and $Z^C$ for CP even sector,
CP odd sector and charged sector respectively. The relations are given by  
\begin{align}\label{eqn:scalar_mixing_matrices}
  \begin{pmatrix} h_1 \\ h_2 \\ h_3 \\ h_4\end{pmatrix} = Z^E  \begin{pmatrix}{\phi_1^0}^r  \\ {\phi_2^0}^r \\  {{\chi^\prime}^0}^r \\ {\chi^0}^r  \end{pmatrix}, \quad 
 \begin{pmatrix} G_1 \\ G_2 \\ A_1 \\ A_2\end{pmatrix} =  Z^O  \begin{pmatrix}{\phi_1^0}^i  \\ {\phi_2^0}^i \\  {{\chi^\prime}^0}^i \\ {\chi^0}^i  \end{pmatrix}, \quad
 \begin{pmatrix} G_1^+ \\ G_2^+ \\ h_1^+ \\ h_2^+\end{pmatrix} = Z^C  \begin{pmatrix}{\phi_1}^+  \\ {\phi_2}^+ \\  {\chi^\prime}^+ \\ {\chi}^+  \end{pmatrix}.
\end{align}
Here $h_1$, $h_2$, $h_3$, $h_4$ are CP even scalar mass eigenstates, $A_1$ and $A_2$ are CP odd scalar mass eigenstates and $h_1^+$, $h_2^+$ are charged scalar mass eigenstates. 
$G_1$, $G_2$ are neutral Goldstone bosons and $G_1^+$, $G_2^+$ are the charged Goldstone bosons.
As discussed in the previous section, the field ${\phi_2}^0$ will have FCNH interactions while the fields  
${\phi_1}^0$, $\chi^0$ and ${\chi^\prime}^0$ have no FCNH interactions. Therefore, any mass eigenstate formed out of the linear 
combinations of the three fields ${\phi_1}^0$, $\chi^0$ and ${\chi^\prime}^0$ can lie at sub-TeV scale. A typical example of the composition of 
the sub-TeV scalar states can be arranged through the following choice of the parameter values in our model: 
 \begin{align*}
 \{M_2, \, \alpha_1, \, \alpha_2, \, \beta_1, \, \beta_2, \, \lambda_2, \, \lambda_3, \, \lambda_4, \, \rho_3 \} = 0, && \\
 u = 12 \,\, {\rm TeV}, \,\, M_1 = -0.3  \,\, {\rm GeV}, \,\, v_2 = 0, \,\, v_3 =  7  \,\, {\rm GeV}, && \\ 
  \rho_1 = 0.1, \,\, \rho_2 = 1, \,\, \lambda_1 = 0.13,  \,\, \beta_3 = 1.4 \,\, {\rm and} \,\, \alpha_3 = 1. 
 \end{align*} 
 For the above choice of parameters the mixing matrices are given by 
 \begin{align}\label{eqn:scalar_mixing_numerical}
  Z^E = \begingroup \setlength\arraycolsep{5pt}
 \begin{pmatrix}
  0.999 & 0 & \sim10^{-8} & 0.034 \\   
  -0.034 & 0 & \sim10^{-7} & 0.999 \\
  0 & 1 & 0 & 0 \\
  \sim -10^{-8} & 0 & \sim 1 & \sim-10^{-7}
 \end{pmatrix}
 \endgroup,  \quad 
 Z^O =  \begingroup \setlength\arraycolsep{5pt}
 \begin{pmatrix}
  0.999 & 0 & 0.02 & 0.028 \\ 
     -0.02 & 0 &  0.999 & 0 \\
  -0.028 & 0 & \sim -10^{-4} & 0.999 \\
  0 & 1 & 0 & 0
 \end{pmatrix}, 
 \endgroup  \\
 Z^C = \begingroup \setlength\arraycolsep{5pt}
 \begin{pmatrix}
  0.999 & 0 & 0 & 0.028 \\   
     0 & 0.02 &  0.999 & 0\\
  -0.028 & 0 & 0 & 0.999 \\
  0 & 0.999 & -0.02 & 0 
 \end{pmatrix}.
 \endgroup \nn \hspace{4cm}
 \end{align}
 \vspace*{0.2cm}
 
The mass eigenvalues are : $m_{h_1} \approx 125$ GeV, $m_{h_2} \approx 300$ GeV, $m_{h_3} \approx 8.5$ TeV, $m_{h_4} \approx 17$ TeV,
$m_{A_1} \simeq 300$ GeV, $m_{A_2}  \simeq 8.5$ TeV, $m_{h_1^+} \simeq 363$ GeV and $m_{h_2^+} \simeq 8.5$ TeV.  We shall keep this 
spectrum for the scalars as our choice for the collider analysis presented later.

The mixing matrices clearly show that the neutral scalars $h_3$ and $A_2$, which are basically ${\phi_2^0}^r$ and ${\phi_2^0}^i$ 
respectively, have been kept unmixed with other neutral scalars, because both ${\phi_2^0}^r$ and ${\phi_2^0}^i$ have FCNH interactions as discussed earlier.

For the choice $u = 12$ TeV and for $g_2 = g_1$, the mass values for $W^\prime$ and $Z^\prime$ are $M_{W^\prime} \simeq 4$ TeV and $M_{Z^\prime} \simeq 4.7$ TeV respectively, and both of them satisfies the current lower 
bounds obtained by the ATLAS collaboration \cite{Aaboud:2017yvp, Aaboud:2017buh}. Also the $Z-Z^\prime$ mixing angle is small 
($\theta_{zz^\prime}\simeq 10^{-4}$) and satisfies the constraint from the electroweak precision data \cite{Erler:2009jh}.
Note that for $v_2 = 0$, the $W-W^\prime$ mixing angle $\theta_{\text{ww}^\prime}$ (see Eq.~\ref{eqn:wwp_mixing_angle}) is zero.

\begin{figure}
 \includegraphics{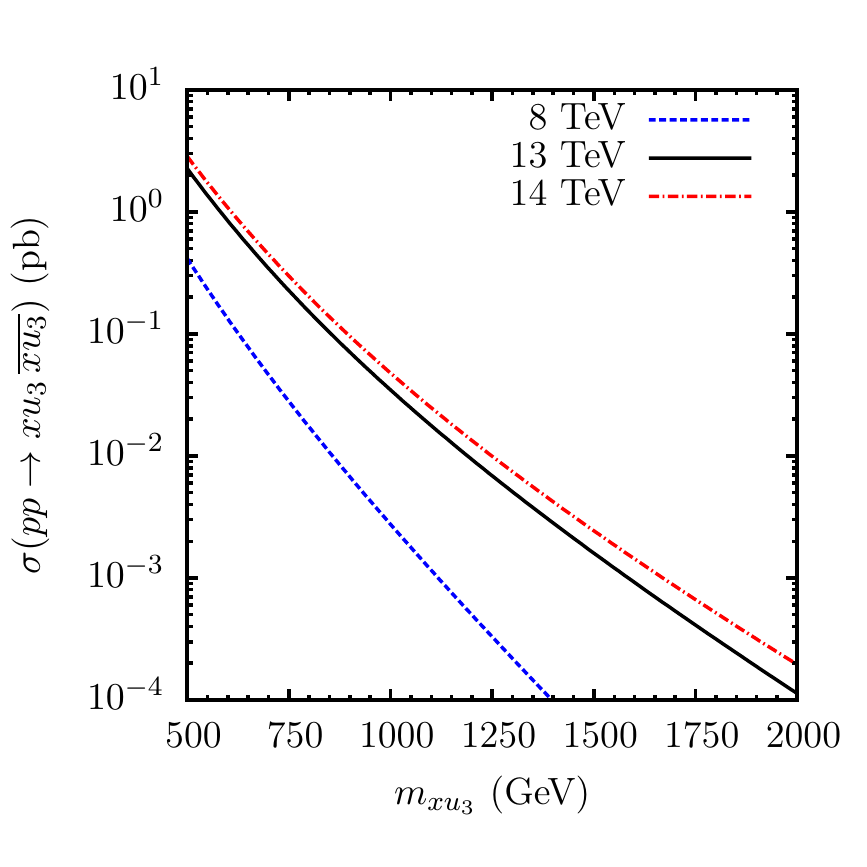}
 \caption{Production cross secttion for the VLQ $xu_3$ for different masses for 8 TeV, 13 TeV and 14 TeV LHC.}
 \label{fig:vlq_production}
\end{figure}
\subsection{Production and decay of the Vector Like Quarks}

As color triplets, the VLQs will be pair-produced at the LHC mostly through strong interactions. 
The pair production cross section at LHC with different center-of-mass energies  and as a function of the mass of $xu_3$ has been 
shown in fig.\ref{fig:vlq_production}. We have used \texttt{NN23LO1}\cite{Ball:2014uwa} parton distribution function with default factorization 
and renormalization scale in \texttt{Madgraph5\_aMC@NLO} for our estimates.

For a given set of allowed values for the parameters in the model, the six VLQs (three up-type and three down-type)  will 
mostly have different signatures depending on the generation they belong to. To discuss the phenomenology and for simplicity we consider the third generation up-type VLQ $xu_3$ to be the lightest one among all VLQs in the model.
Since we have already discussed in detail the mixing between VLQs and SM quarks,  we list the relevant interaction terms 
for $xu_3$ in table \ref{table:vlq_interactions}. As the interactions with the physical scalar fields would look quite cumbersome and 
messy, we have chosen to show the interaction of the physical (mass eigenstates) fermions with the scalars in the gauge eigenbasis.
The interaction terms with the physical scalars can be obtained by using the rotated fields in terms of the elements of $4\times 4$ matrices $Z^E$, $Z^O$ and $Z^C$.  For example, the interaction term for $A_1$ will be 
\begin{align*}
 A_1 \, \bar{t} \, xu_3 :& \,\,  \Big[(Z^O)^T_{13} \, K_{{\phi_1^0}^i} (C^S_{{\phi_1^0}^i} + C^P_{{\phi_1^0}^i} \gamma^5) + 
 (Z^O)^T_{23} \, K_{{\phi_2^0}^i} (C^S_{{\phi_2^0}^i} + C^P_{{\phi_2^0}^i} \gamma^5)  \\ 
 & +  (Z^O)^T_{33} \, K_{{{\chi^\prime}^0}^i} (C^S_{{{\chi^\prime}^0}^i} + C^P_{{{\chi^\prime}^0}^i} \gamma^5) +
  (Z^O)^T_{43} \, K_{{{\chi}^0}^i} (C^S_{{{\chi}^0}^i} + C^P_{{{\chi}^0}^i} \gamma^5) \Big]  \,.
\end{align*}

\begin{table}
 \begin{tabular}{|c|c|c|c|}
 \hline
    \boldmath{${\phi} \, \bar{t} \, xu_3$}    & \boldmath{$K_{\phi}$} & \boldmath{$C^S_{\phi}$} & \boldmath{$C^P_\phi$} \\ \hline
  \boldmath{${\phi_1^0}^r \, \bar{t} \, xu_3$}        & $-\frac{1}{2\,v_1} (c_L^t m_t c_R^t + s_L^t m_{xu_3} s_R^t)$ & $c_L^t s_R^t + s_L^t c_R^t$ & $c_L^t s_R^t - s_L^t c_R^t$ \\ \hline
  {\multirow{3}{*}{\boldmath{${\phi_2^0}^r \, \bar{t} \, xu_3$}}}& {\multirow{3}{*}{$-\frac{1}{2\,v_1} (c_L^b m_b c_R^b + s_L^b m_{xd_3} s_R^b)$}} & $(V_L^{CKM})_{tb} \frac{s_R^t}{c_L^b}$                  & $(V_L^{CKM})_{tb} \frac{s_R^t}{c_L^b}$ \\ 
                                                                 &                                                                                 &                           $+$                           &         $-$                            \\
                                                                 &                                                                                 & $({V_L^{CKM}}^*)_{tb} \frac{c_R^t s_L^t}{c_L^b c_L^t}$  & $({V_L^{CKM}}^*)_{tb} \frac{c_R^t s_L^t}{c_L^b c_L^t}$ \\ \hline
  \boldmath{${\chi^0}^r \, \bar{t} \, xu_3$}          & $\frac{1}{2\,v_3} (c_L^t m_t s_R^t - s_L^t m_{xu_3} c_R^t)$  & $c_L^t c_R^t - s_L^t s_R^t$ & $c_L^t c_R^t + s_L^t s_R^t$ \\ \hline
  \boldmath{${{\chi^\prime}^0}^r \, \bar{t} \, xu_3$} & $\frac{1}{2\,u} (s_L^t m_t s_R^t + c_L^t m_{xu_3} c_R^t)$    & $s_L^t c_R^t + c_L^t s_R^t$ & $s_L^t c_R^t - c_L^t s_R^t$ \\ \hline
  \boldmath{${\phi_1^0}^i \, \bar{t} \, xu_3$}        & $-\frac{i}{2\,v_1} (c_L^t m_t c_R^t + s_L^t m_{xu_3} s_R^t)$ & $c_L^t s_R^t - s_L^t c_R^t$ & $c_L^t s_R^t + s_L^t c_R^t$ \\ \hline
    {\multirow{3}{*}{\boldmath{${\phi_2^0}^i \, \bar{t} \, xu_3$}}}& {\multirow{3}{*}{$\frac{i}{2\,v_1} (c_L^b m_b c_R^b + s_L^b m_{xd_3} s_R^b)$}} & $(V_L^{CKM})_{tb} \frac{s_R^t}{c_L^b}$                  & $(V_L^{CKM})_{tb} \frac{s_R^t}{c_L^b}$ \\ 
                                                                 &                                                                                 &                           $-$                           &         $+$                            \\
                                                                 &                                                                                 & $({V_L^{CKM}}^*)_{tb} \frac{c_R^t s_L^t}{c_L^b c_L^t}$  & $({V_L^{CKM}}^*)_{tb} \frac{c_R^t s_L^t}{c_L^b c_L^t}$ \\ \hline
  \boldmath{${\chi^0}^i \, \bar{t} \, xu_3$}          & $\frac{i}{2\,v_3} (c_L^t m_t s_R^t - s_L^t m_{xu_3} c_R^t)$  & $c_L^t c_R^t + s_L^t s_R^t$ & $c_L^t c_R^t - s_L^t s_R^t$ \\ \hline
  \boldmath{${{\chi^\prime}^0}^i \, \bar{t} \, xu_3$} & $\frac{i}{2\,u} (s_L^t m_t s_R^t + c_L^t m_{xu_3} c_R^t)$    & $s_L^t c_R^t - c_L^t s_R^t$ & $s_L^t c_R^t + c_L^t s_R^t$ \\
  \hline
  {\multirow{3}{*}{\boldmath{${\phi_1^-} \, \bar{b} \, xu_3$}}} & {\multirow{3}{*}{$\frac{1}{\sqrt{2}\, v_1}$}} &$-s_L^t c_R^b (c_L^t m_t c_R^t + s_L^t m_{xu_3} s_R^t)$ & $s_L^t c_R^b (c_L^t m_t c_R^t + s_L^t m_{xu_3} s_R^t)$\\
                                                                &                                               &                           $+$                          &                            $+$                        \\
                                                                &                                               &$c_L^b s_R^t (c_L^b m_b c_R^b + s_L^b m_{xd_3} s_R^b)$  & $c_L^b s_R^t (c_L^b m_b c_R^b + s_L^b m_{xd_3} s_R^b)$\\                                                                                                               
 \hline
   {\multirow{3}{*}{\boldmath{${\phi_2^-} \, \bar{b} \, xu_3$}}} & {\multirow{3}{*}{$\frac{({V_L^{CKM}}^*)_{tb}}{\sqrt{2}\, v_1}$}} &$-\frac{s_R^t}{c_L^t} (c_L^t m_t c_R^t + s_L^t m_{xu_3} s_R^t)$ & $-\frac{s_R^t}{c_L^t} (c_L^t m_t c_R^t + s_L^t m_{xu_3} s_R^t)$\\
                                                                &                                               &                           $+$                          &                            $-$                        \\
                                                                &                                               &$\frac{s_L^t\,c_R^b}{c_L^t\, c_L^b}(c_L^b m_b c_R^b + s_L^b m_{xd_3} s_R^b)$  & $\frac{s_L^t\,c_R^b}{c_L^t\, c_L^b}(c_L^b m_b c_R^b + s_L^b m_{xd_3} s_R^b)$ \\                                                                                                               
 \hline
 {\multirow{3}{*}{\boldmath{${\chi^-} \, \bar{b} \, xu_3$}}}    & {\multirow{3}{*}{$\frac{({V_L^{CKM}}^*)_{tb}}{\sqrt{2}\, v_3}$}} & $\frac{c_R^t}{c_L^t}(c_L^t m_t s_R^t - s_L^t m_{xu_3} c_R^t)$                &      $\frac{c_R^t}{c_L^t}(c_L^t m_t s_R^t - s_L^t m_{xu_3} c_R^t)$    \\
                                                                &                                                                  &                  $+$                                                         &                            $-$                                          \\
                                                                &                                                                  & $\frac{s_R^b\, s_L^t}{c_L^b \,c_L^t}(c_L^b m_b s_R^b - s_L^b m_{xd_3} c_R^b)$& $\frac{s_R^b\, s_L^t}{c_L^b \,c_L^t}(c_L^b m_b s_R^b - s_L^b m_{xd_3} c_R^b)$\\
 \hline
   {\multirow{3}{*}{\boldmath{${{\chi^\prime}^-} \, \bar{b} \, xu_3$}}}  & {\multirow{3}{*}{$\frac{1}{\sqrt{2}\, u}$}} &$s_L^b c_R^t (s_L^t m_t s_R^t + c_L^t m_{xu_3} c_R^t)$ & $s_L^b c_R^t (s_L^t m_t s_R^t + c_L^t m_{xu_3} c_R^t)$ \\
                                                                &                                               &                           $-$                          &                            $+$                        \\
                                                                &                                               &$c_L^t s_R^b (s_L^b m_b s_R^b + c_L^b m_{xd_3} c_R^b)$  & $c_L^t s_R^b (s_L^b m_b s_R^b + c_L^b m_{xd_3} c_R^b)$\\ 
\hline \hline
    \boldmath{$a_\mu \bar{t} \, xu_3 $}          & \boldmath{$K_a$} & \boldmath{$C^V_a$} & \boldmath{$C^A_a$} \\ \hline
 \boldmath{$Z_\mu \bar{t} \, xu_3 $} & $\frac{g}{4 \cos{\theta_W}}$         & $-s_L^t c_L^t$ & $s_L^t c_L^t$ \\ \hline
  \boldmath{$W_\mu^{-} \bar{b} \, xu_3 $} & $\frac{g}{2\sqrt{2}}({V_L^{CKM}}^*)_{tb}$         & $\frac{-s_L^t}{c_L^t}$ & $\frac{s_L^t}{c_L^t}$ \\ \hline
 \end{tabular}
 \caption{The interaction terms including scalars are of the form $\phi \, K_\phi(C^S_\phi + C^P_\phi \gamma^5)\, xu_3$ and including gauge bosons are of the form $ a_\mu \, \bar{t} \, K_a\gamma^\mu(C^V_a + C^A_a \,  \gamma^5) \, xu_3$. The interactions 
 for the physical scalars can be obtained using the transformations given by Eq. \ref{eqn:scalar_mixing_matrices}.}
 \label{table:vlq_interactions}
\end{table}

\begin{figure}[t]
 \includegraphics[height=4in,width=4in]{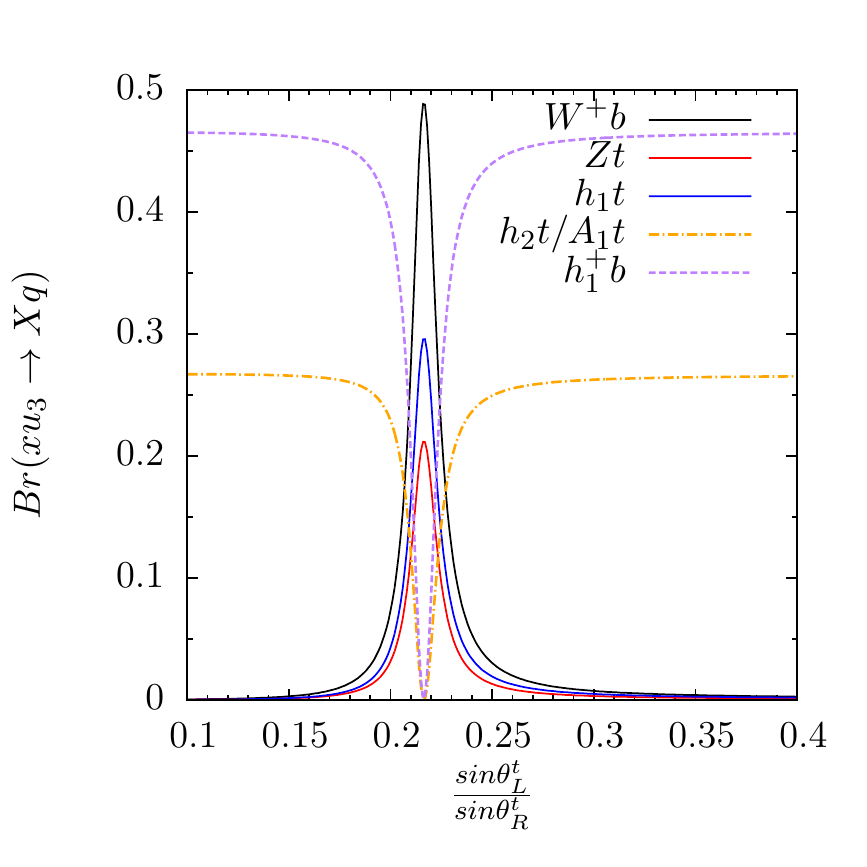}
 \caption{Branching ratios for different decay modes of the VLQ $xu_3$ as a function of mixing angle $\theta_L^t$. 
          For the plot $m_{xu_3} = 800$ GeV, $m_{h_2} = m_{A_1} = 300$ GeV, $m_{h_1^+} = 363$ GeV, $m_{xd_3} = 5$ TeV, 
          $\sin\theta_L^b =10^{-4}$, $\sin\theta_R^b \simeq 10^{-4}$ and $\sin\theta_R^t = 10^{-3}$.}
 \label{fig:vlq_branchings}
\end{figure}

The possible final states that $xu_3$ can decay to are $t\, Z$, $t \,h_1$, $b \,W^+$, $t \,h_2$, $t \,A_1$ and  $b \, h_1^+$, since the scalars $h_3$, $h_4$, $A_2$, $h_2^+$ and the new gauge bosons $Z^\prime$, $W^\prime$ are heavier compared to the VLQ $xu_3$. 
For small mixing angles the ``non-standard" decay modes ($t \,h_2$, $t \,A_1$, $b \, h_1^+$) will mostly dominate over the standard decay modes ($t\, Z$, $t \,h_1$, $b \,W^+$) because of 
the presence of direct Yukawa interaction term,  $Y^{qxu}$ in the Lagrangian.
The standard decay modes will start dominating once  $Y^{qxu}$ tends to zero. This feature is illustrated in  
Fig. \ref{fig:vlq_branchings} which shows
the branching ratios for different decay modes for a 800 GeV $xu_3$ as function of mixing angles, where we have fixed 
$\sin\theta_R^t = 10^{-3}$  and varied $\sin\theta_L^t$ accordingly.  
 
We have considered small mixing angles ($\simeq 10^{-3}$) to avoid constraints from flavour sector and electroweak precision 
data \cite{AguilarSaavedra:2002kr, Cacciapaglia:2011fx, Aguilar-Saavedra:2013qpa, 
Aguilar-Saavedra:2013wba, Alok:2014yua, Alok:2015iha}  and as an example, we have checked that the contribution
of VLQs to the $K-\bar{K}$ oscillation parameter $\Delta m_K$ is few orders of magnitude less compared to the SM value.  We find that the non-standard decay modes dominate the standard decay 
modes except where $\dfrac{\sin\theta_L^t}{\sin\theta_R^t}$ lies in the small range 0.2-0.23. We can understand this feature of the 
decay probability by looking at the interaction terms 
${\chi^0}^r \, \bar{t} \, xu_3$, ${\chi^0}^i \, \bar{t} \, xu_3$ and ${\chi^-} \, \bar{b} \, xu_3$ from Table \ref{table:vlq_interactions}. 
In the limit $\dfrac{\tan\theta_L^t}{\tan\theta_R^t} = \dfrac{m_t}{m_{xu_3}}$, the coupling strengths for the interactions 
${\chi^0}^r \, \bar{t} \, xu_3$ and ${\chi^0}^i \, \bar{t} \, xu_3$ is identically zero. 
Moreover, the same limit along with small values of $\sin\theta_L^t$ and $\sin\theta_R^b$ make the coupling strength for the interaction 
${\chi^-} \, \bar{b} \, xu_3$ 
negligibly small. Thus when the ratio of mixing angles become 
\begin{align}
 \frac{\sin\theta_L^t}{\sin\theta_R^t} \simeq \frac{\tan\theta_L^t}{\tan\theta_R^t} = \frac{m_t}{m_{xu_3}} \simeq \frac{173}{800} \simeq 0.216,
\end{align}
the interactions with the scalars $h_2$, $A_1$ and $h_1^+$ goes to zero. Consequently the decays of $xu_3$ to the 
SM particles enhances. 
In the next section we study the possible collider signatures for the scenario where the branching ratios for the VLQ $xu_3$ lie
away from the standard mode dominated region such that after production  $xu_3$ decays to one of the final states from $t \,h_2$, $t \,A_1$, $b \, h_1^+$.

The collider signatures of $xu_3$ will eventually depend on the decay modes of the the scalars $h_2$, $A_1$ and $h_1^+$. 
From Eq.~\ref{eqn:scalar_mixing_matrices} and Eq.~\ref{eqn:scalar_mixing_numerical} we note that the scalar 
$h_2$ is made up of a very small component ($\sim10^{-2}$) of one of the real neutral part of the bi-doublet Higgs field (${\phi_1^0}^r$) and 
a large component of the real neutral part of $H_1$, i.e., ${\chi^0}^r$. From the Yukawa terms in the
Lagrangian it can be seen that $H_1$ gives mass to the charged leptons but there is no Yukawa 
interaction term involving SM quarks and $H_1$. Hence the strength of the Yukawa interaction for the 
mass eigenstate $h_2$ with the SM quarks is negligible compared to the coupling strength with the leptons.
Hence $h_2$ will mostly decay to leptons compared to the SM quarks. The same argument is also applicable 
for $A_1$ and $h_1^+$, because all of them are largely composed of $H_1$.
Note that the charged leptons get masses from the VEV of the $SU(2)_L$ doublet scalar $H_1$ 
which gets a small VEV, $\dfrac{v_3}{\sqrt{2}} \sim 5$ GeV. 
The Yukawa coupling strengths for the scalars $h_2, \, A_1, \, h_1^+$ 
with the leptons from different generations follow the mass hierarchy
and with more than $99\%$ probability the scalar $h_2$ and $A_1$  will decay to $\tau^+\,\tau^-$ whereas 
$h_1^+$ will decay to $\tau^+\,\nu_\tau$. 
\subsection{Benchmark points}
For the collider analysis we have chosen three benchmark points based on $xu_3$ mass. The pair production 
cross section and the branching ratios for different benchmarks are given in table~\ref{table:benchmarks}. 
For all the three cases the masses for the scalars $h_2$, $A_1$ and $h_1^+$ are kept fixed at 300 GeV, 300 GeV and 363 GeV respectively.
Due to the smallness of the Yukawa couplings with the SM quarks, the scalars $h_2$ and $A_2$ can not be produced efficiently
at the LHC via gluon fusion and the production cross sections of them are few tens of fb  for 13 TeV center of mass energy.  
Hence, the experimental limits on on their masses are fairly weak. Note that the Yukawa couplings of these scalars 
with the VLQs are also very small and the VLQ loops contribute very less towards their production. The scenario is almost same as the lepton-specific two-Higgs doublet 
model where the limit on the massive states are of the order of 180-200 GeV \cite{Abdallah:2004wy}.
Since the production cross-section falls off rapidly for higher masses we have used an 
integrated luminosity of 100 fb$^{-1}$ for BP1 whereas 3000 fb$^{-1}$ luminosity is used for the analysis of BP2 and BP3. 

 We would like to emphasise that the benchmark points chosen here are fairly general as long as the mass of $xu_3$ is greater than the masses of the scalars $h_2$, $A_1$ and $h_1^+$,
such that the non-standard decay modes are kinematically allowed. From figure \ref{fig:vlq_branchings} it can be observed 
that the decay branchings depend mildly on the ratio of the mixing angles $\theta_L$ and $\theta_R$, once we are away from the narrow peak region. 
Also the decay branchings of $h_2$, $A_1$ and $h_1^+$ to tau leptons depend on the yukawa couplings and are almost independent of the masses of the scalars.
 \begin{table}[h]
  \begin{tabular}{|c|c|c|c|c|c|}
   \hline
   Benchmarks & $m_{xu_3}$ & Br($xu_3 \rightarrow h_2 t$)& Br($xu_3 \rightarrow A_1 t$)& Br($xu_3 \rightarrow h_1^+ b$) &   $\sigma (p p \rightarrow xu_3, \ \overline{xu_3})$ \\
   \hline
   BP1 & 1 TeV& 0.26 & 0.26 & 0.48 & 32.33 fb \\
   \hline
   BP2 &1.5 TeV & 0.255 & 0.255 &0.49 & 1.554 fb\\
   \hline
   BP3 &2 TeV &0.25 & 0.25& 0.5& 0.113 fb \\
   \hline
  \end{tabular}
\caption{Different benchmark scenarios we have used in collider analysis. }
\label{table:benchmarks}
 \end{table}

\section{Collider Analysis}\label{section:collider}
\begin{figure}
 \includegraphics[width=12cm]{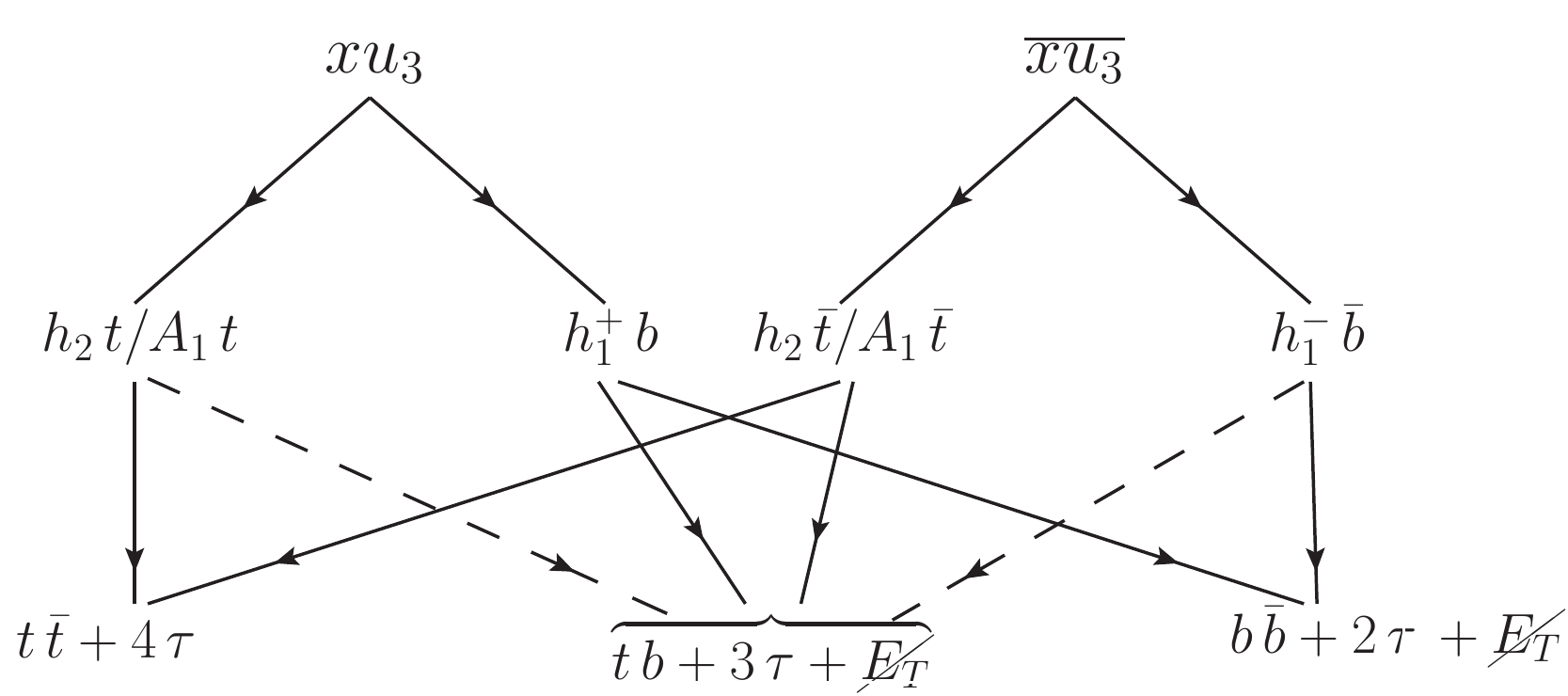}
 \caption{All possible final states resulting from the pair production of $xu_3$ and their subsequent non-standard decay.}
 \label{fig:finalstates}
\end{figure}

Now we consider the collider signatures of $xu_3$ for 13 TeV LHC in the scenario where the pair produced
$xu_3$ will decay to the final states $t \,h_2$, $t \,A_1$, $b \, h_1^+$  and the scalars $h_2$, $A_1$ and $h_1^+$ will further decay to tau leptons. 
Taking into account all possible decay chains of $xu_3$, fig.\ref{fig:finalstates} shows 
all possible final states that can arise from the pair production of $xu_3$. Since each final state contains at least 
two tau leptons and at least one $b$ quark, we choose the final state  with
at least two $\tau$-tagged jets,  at least three non $\tau$-tagged jets among which at least one is b-tagged,  
and at least one lepton ($\geq 3j(1b) \, + \geq 2\tau \,  + \geq 1l$) for the collider analysis. 

The possible SM processes that can contribute as background to the above choice of final state are the following:
\begin{itemize}
 \item $p p \rightarrow t \bar{t} \, \, + \text{jets}$,
 \item $p p \rightarrow t \bar{t} \, l^+ l^- + \text{jets}$ \,\,\, and \, $p p \rightarrow t \bar{t} \, \tau^+ \tau^- + \text{jets}$,
 \item $p p \rightarrow t \bar{t} \, l^+ \nu_l \, + \text{jets}$ \,\, and \, $p p \rightarrow t \bar{t} \, \tau^+ \nu_\tau \, + \text{jets}$,
 \item $p p \rightarrow Z Z Z$,
 \item $p p \rightarrow W^\pm/Z + \text{jets}$.
\end{itemize}
Among all the above possible backgrounds the most dominant one is $t\bar{t}+\text{jets}$. Although the cross section
for the backgrounds $W^\pm/Z + \text{jets}$ is large it is possible to get rid of this by a large $\slashed{E_T}$ requirement which we have used in our analysis. The contribution of $ZZZ$ will be negligible because of its small cross section.
Hence we consider only the first three of the above list of backgrounds for the collider analysis in the context of 13 TeV LHC.

To study the collider phenomenology we implemented the model in the spectrum-generator-generator {\tt SARAH} \cite{Staub:2013tta}. The source code generated by {\tt SARAH} for the spectrum 
generator {\tt SPheno} \cite{Porod:2011nf} has been used in {\tt SPheno} to study the spectrum of the model. The files generated by {\tt SARAH} in the {\tt UFO} format and the spectrum file 
generated by {\tt SPheno} has been used in {\tt MadGraph 5} \cite{Alwall:2014hca} for event generation for the signal. The background events have also been generated using {\tt Madgraph 5}.
For showering and hadronization we used {\tt Pythia 6} \cite{Sjostrand:2006za} interfaced in {\tt Madgraph 5}. {\tt DELPHES 3} \cite{deFavereau:2013fsa} within CMS environment 
has been used to take into account the detector effects and also for reconstruction of the final state objects. The anti-$k_T$ algorithm  with cone size 0.5 have been used for the jet reconstruction. For the reconstruction of the jets, {\tt FastJet} \cite{Cacciari:2011ma} embedded in {\tt DELPHES} has been used. 
{\tt MadAnalysis 5} \cite{Conte:2012fm} package has been used for the event-analysis using the event format {\tt ROOT} and {\tt LHCO}. 

The selection criteria for the final state objects in the reconstructed events are such that a non $\tau$-tagged jet with $p_T(j)>20$ GeV and $|\eta(j)|<2.5$ is considered
in the event, an electron or a muon with  $p_T(l)>10$ GeV and $|\eta(l)|<2.5$ are considered in the event, a $\tau$-tagged jet with $p_T(\tau)>20$ GeV and $|\eta(\tau)|<2.5$
is considered in the event. Note that here $j$  denotes a non $\tau$-tagged jet, $\tau$ denotes a $\tau$-tagged jet. 
A non $\tau$-tagged jet ($j$) is either a light jet or a $b$-tagged jet.  The minimum angular separation between all final state objects satisfy $\Delta R > 0.4$. The $\tau$-tagging and mistagging efficiencies are incorporated in \texttt{Delphes3}  as reported by the ATLAS collaboration~\cite{ATL-PHYS-PUB-2015-045}. We operate our 
simulation on the Medium tag point for which the tagging efficiency of 1-prong (3-prong) $\tau$ decay is 70\% (60\%)
and the corresponding mistagging rate is 1\% (2\%).  
 \begin{figure}[t]
 \includegraphics[width=.49\linewidth,height=2.1in]{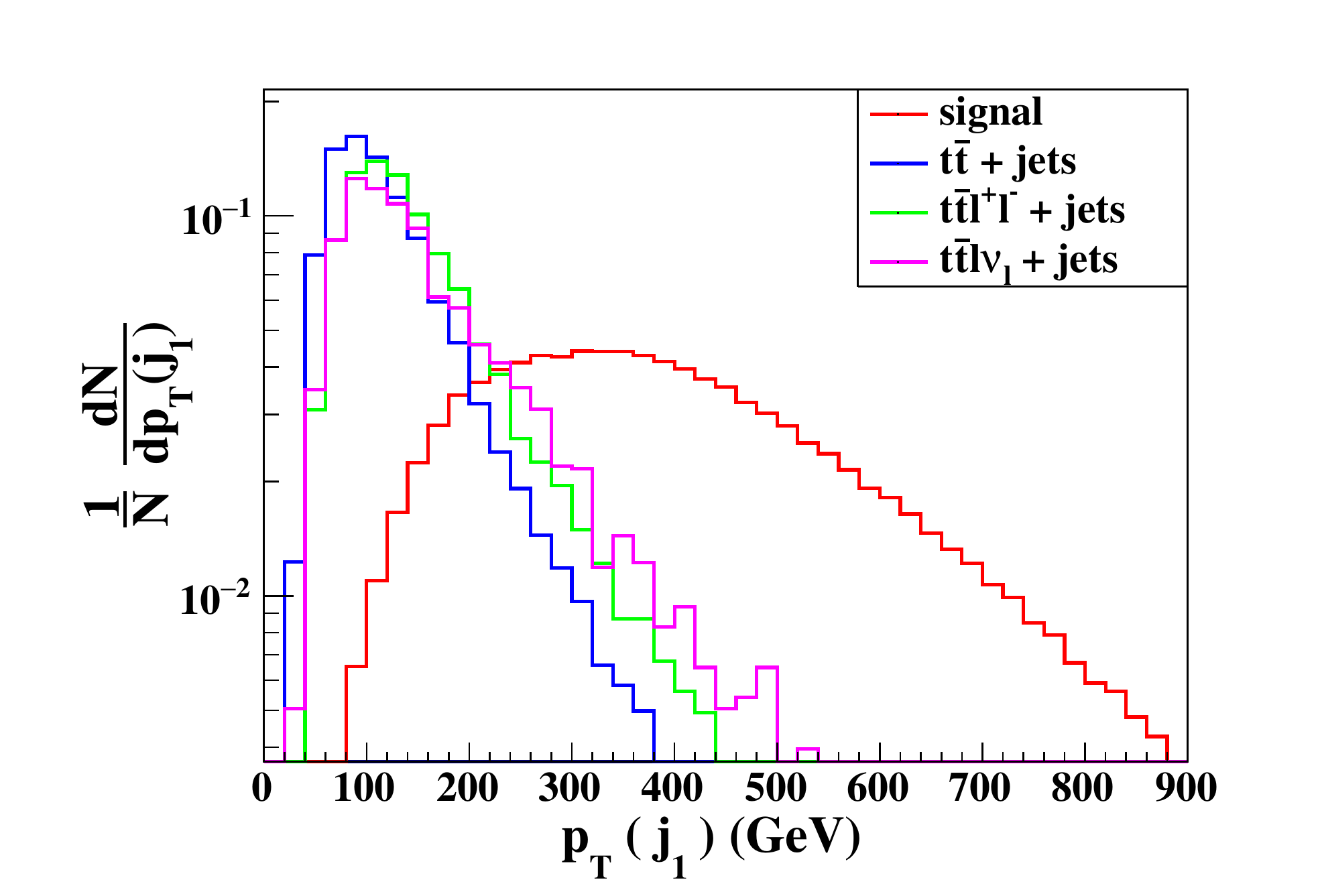}
 \hfill
 \includegraphics[width=0.5\linewidth,height=2.1in]{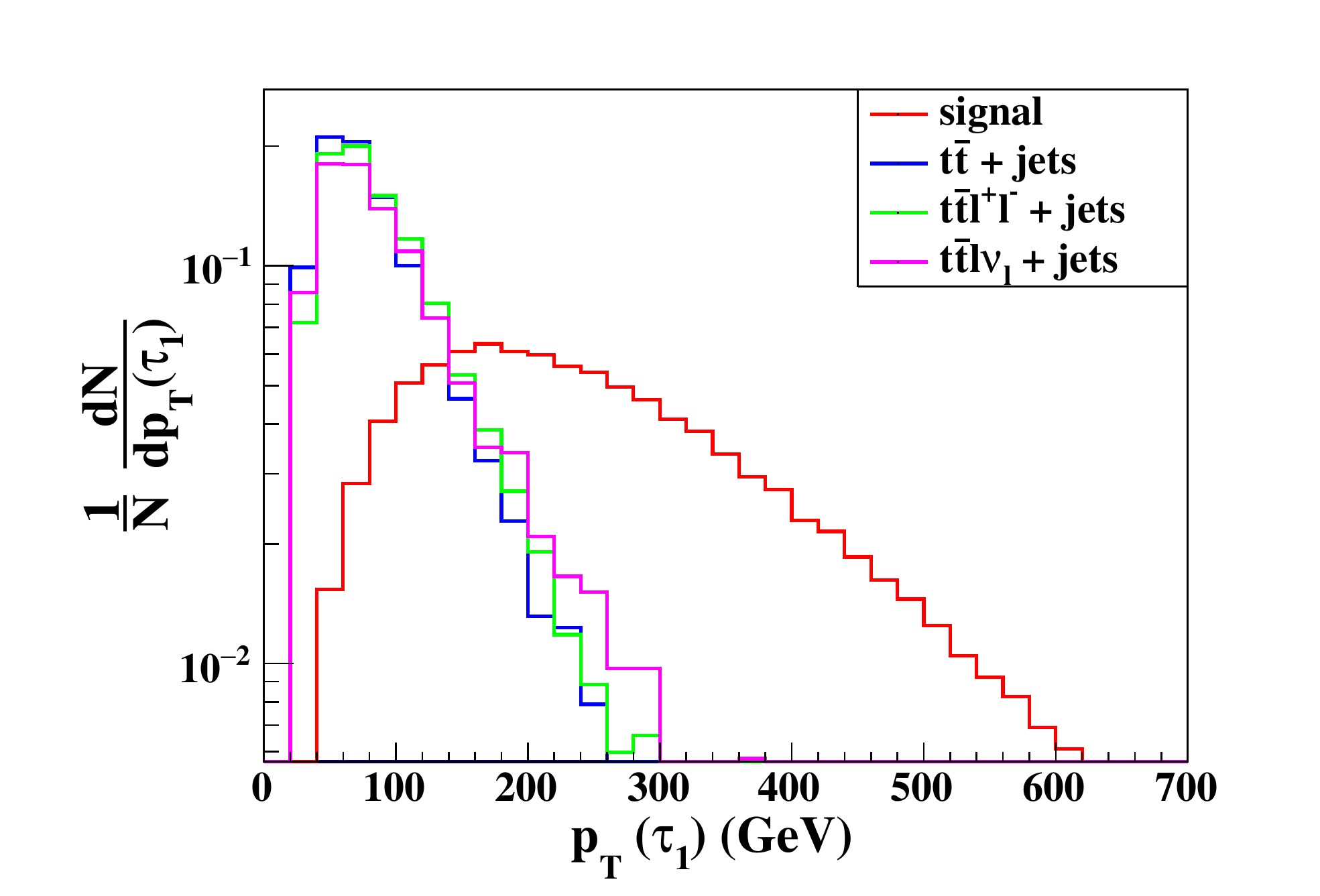}
 \includegraphics[width=0.5\linewidth,height=2.1in]{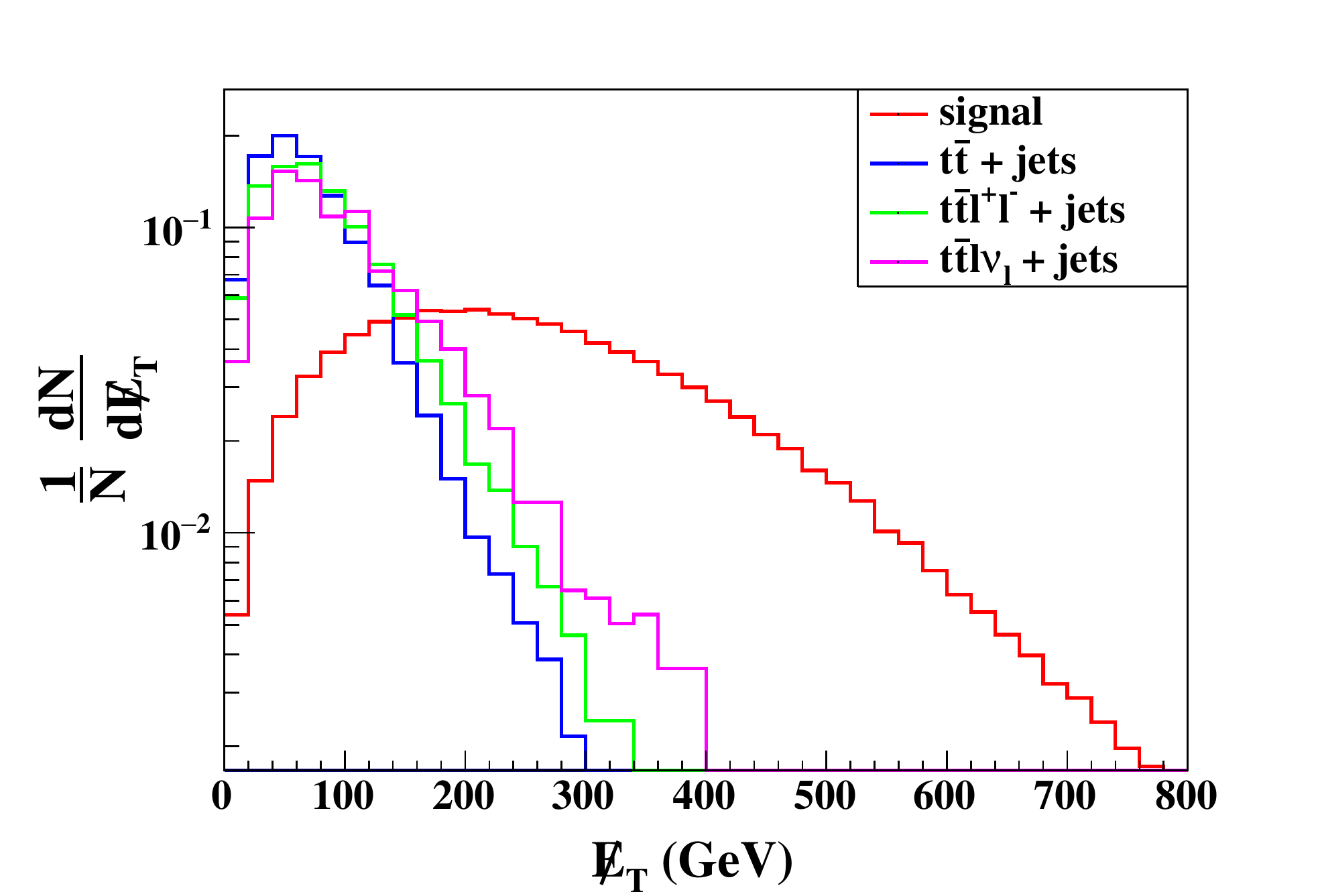}
 \caption{Normalized distributions for the transverse momentum for the leading non $\tau$-tagged jet ($P_T(j_1)$),
 the transverse momentum for the leading $\tau$-tagged jet ($P_T(\tau_1)$) and the total missing transverse energy $\slashed{E_T}$ for BP1 with $m_{xu_3} = 1$ TeV.}
 \label{fig:1000GeV_analysis}
 \end{figure}
 
 \subsection{BP1 : \boldmath{$m_{xu_3} = 1$} TeV with $100\, \text{fb}^{-1}$}
 For BP1 the production cross-section and different branching ratios are tabulated in table~\ref{table:benchmarks}.
 For the background simulation we generated $p p \rightarrow t \bar{t}$ events up to two additional jets at the
 leading order accuracy  and used {\it shower}-$K_T$ matching  scheme in {\tt Madgraph 5} to avoid the double counting
 between the partonic events and showered events. For the event analysis we used the cross section for 13 TeV at 
 the NNLO accuracy for top quark pair production, i.e., 815.96 pb\cite{Czakon:2011xx}. By following the same procedure we have generated both the $t\bar{t}l^{+}l^{-} $ and  $t\bar{t}l \nu_l $ events up to two additional jets at the leading order accuracy. 
  The obtained leading order cross section at the parton level for $t\bar{t}l^{+}l^{-} + \leq 2\,\text{jets}$ is 96 fb for 13 TeV LHC and to accommodate the NLO effects we multiplied the cross section with a factor
  of 1.4 which is the NLO K-factor for $t\bar{t}Z$. Similarly for $t\bar{t}l \nu_l + \leq \, 2\,\text{jets}$ 
 we multiply the leading order parton level cross section 166 fb for 13 TeV LHC with 1.4 which is the K-factor for $t\bar{t}W$\cite{Alwall:2014hca}. 


 After the event selection we compare the phase space behavior of the signal events with the background and plot 
 the normalized distributions of the transverse momentum ($p_T$) of the leading non $\tau$-tagged jet and leading $\tau$-tagged 
 jet along with ($\slashed{E_T}$) in fig.~\ref{fig:1000GeV_analysis}. Due to the large mass separation between the VLQ and the 
 scalars, the leading jet is expected to be quite hard as the figure shows. In addition, the $\tau$ which comes from the decay 
 of the scalar which is around 300 GeV in mass is also quite hard in $p_T$.  Thus one can make a quite easy separation of the 
 signal and background events using the distributions shown in ~\ref{fig:1000GeV_analysis}.  Based on the distributions and 
 to further optimize the  signal-to-background ratio, we apply the following kinematic cuts on the final state objects:
\begin{equation}
 p_T(j_1) \geq 200 \, \text{GeV}, \, \, \, \, p_T(\tau_1) \geq 150 \, \text{GeV}, \,\,\,\, \slashed{E_T} \geq  150 \, \text{GeV}.
\end{equation}
With an integrated luminosity of 100 fb$^{-1}$ the cut flow for the signal and background events is shown in table~\ref{table:BP1}.
With the above cuts and with $100$ fb$^{-1}$ integrated luminosity, the statistical significance for BP1 is quite large 
($\sim 12\sigma$).  We have used $\sqrt{2((s+b)\,ln(1+\frac{s}{b})-s)}$ to calculate the significance. Thus BP1 seems to be have VLQs 
in a mass range which would be very close to the current sensitivity of the LHC run if such a final state is analyzed for VLQs
decaying in non-standard channels as in our model.

\subsection{BP2 : \boldmath{$m_{xu_3} = 1.5$} TeV with $ 3000\, \text{fb}^{-1}$}

For BP2 the branching ratios of $xu_3$ for the decay modes $h_2 \, t$, $A_2 \, t$ and $h_2^+ \, t$ are around 
$25.5\%$, $25.5\%$ and $49\%$ respectively.
The background events and their corresponding cross sections are same as in case of BP1 and we have used the 
same preselection criteria on the events (i.e. $\geq 3j(1b) \, + \geq 2\tau \,  + \geq 1l$) for BP2. 
The differences between the mass of the VLQ $xu_3$ and the masses of scalars ($h_2, \, A_1, \,h_1^+$) increase
as we go higher in values of the mass of $xu_3$ while keeping the mass of the scalars fixed as before. It is worth pointing out here
that even if the mass of the scalars are made larger, the decay probabilities of the VLQ do not change much. Therefore the 
event rates would remain the same, albeit the cut efficiencies would change due to new thresholds for the leading jet and tagged $\tau$. 
Note that the $b$ quark that will originate from the decay $xu_3 \rightarrow h_1^+ \, b$ for a 1.5 TeV $xu_3$
will have a large $p_T$ compared to the $b$ quark that originates from the decay of a 1 TeV $xu_3$.
Since the leading non $\tau$-tagged jet is most likely the b jet coming from $h_1^+ \, b$ mode it will be in general 
harder in BP2 compared to BP1.
Accordingly for BP2, we have applied the following selection criteria on the final state objects from the reconstructed events 
to optimize the significance :
\begin{equation}
 p_T(j_1) \geq 300 \, \text{GeV}, \, \, \, \, p_T(\tau_1) \geq 200 \, \text{GeV}, \,\,\,\, \slashed{E_T} \geq 200\, \text{GeV}.
\end{equation}
Notice after the cut on $p_T(j_1)$ to further improve the significance we have also applied cuts with higher values on $p_T(\tau_1)$ and $\slashed{E_T}$ compared to 
the scenario of BP1.
For BP2 with $3000\,\text{fb}^{-1}$ luminosity the cut flow can be found in the table.\ref{table:BP2}.
Using the survived events after the $\slashed{E_T}$ cut we get a significance around $6.2\sigma$ for BP2 with 
the high-luminosity (HL) option at the LHC.

 \begin{table}
 \begin{subtable}{.5\linewidth}
 \centering
 \begin{tabular}{|c|c|c|}
 \hline
 \multicolumn{3}{|c|}{\boldmath{\textbf{BP1} : $m_{xu_3} = 1$ \textbf{TeV}}, \, $\mathcal{L} = 100 \, \text{fb}^{-1}$} \\
 \hline
  {\multirow{2}{*}{Cuts}} & \multicolumn{2}{c|}{No. of Events} \\ \cline{2-3}
                          & Signal    & Background \\        \hline
 Preselection                &     365            &  19677        \\ \hline
$p_T(j_1) \geq 200$ GeV          &   320         &   2959         \\        \hline
$p_T(\tau_1) \geq 150$ GeV          &  245         &    839        \\        \hline
$\slashed{E_T} \geq 150$ GeV          &  188        &    191        \\        \hline
 \end{tabular}
 \caption{}
  \label{table:BP1}
 \end{subtable}
 \begin{subtable}{.4\linewidth}
 \centering
 \begin{tabular}{|c|c|c|}
 \hline
  \multicolumn{3}{|c|}{\textbf{BP2} : \boldmath{$m_{xu_3} = 1.5$ \textbf{TeV}}, \, $\mathcal{L} = 3000 \, \text{fb}^{-1}$} \\
 \hline
  {\multirow{2}{*}{Cuts}} & \multicolumn{2}{c|}{No. of Events} \\ \cline{2-3}
                          & Signal    & Background \\        \hline
 Preselection                &  455      &     590310        \\ \hline
$p_T(j_1) \geq 300$ GeV          & 401        &    28547        \\        \hline
$p_T(\tau_1) \geq 200$ GeV          &   307      &     6050       \\        \hline
$\slashed{E_T} \geq 200$ GeV          &   245       &    1455        \\        \hline
 \end{tabular}
 \caption{}
  \label{table:BP2}
 \end{subtable}
 
 \begin{subtable}{.5\linewidth}
 \centering
 \begin{tabular}{|c|c|c|}
 \hline
   \multicolumn{3}{|c|}{\textbf{BP3} : \boldmath{$m_{xu_3} = 2$ \textbf{TeV}}, \, $\mathcal{L} = 3000 \, \text{fb}^{-1}$} \\
 \hline
  {\multirow{2}{*}{Cuts}} & \multicolumn{2}{c|}{No. of Events} \\ \cline{2-3}
                          & Signal    & Background \\        \hline
 Preselection                &  86     &   856721         \\ \hline
$p_T(j_1) \geq 350$ GeV          &   81     &  269121         \\        \hline
$p_T(j_2) \geq 100$ GeV          &   78     &  252922        \\        \hline
$p_T(\tau_1) \geq 150$ GeV          &  58      & 57285         \\        \hline
$\slashed{E_T} \geq 200$ GeV          &   52     &   11571       \\        \hline
$M_{eff} \geq 2.6$ TeV              &   40     &   378      \\        \hline
 \end{tabular}
 \caption{}
 \label{table:BP3}
 \end{subtable}
 \caption{Cut flow table for BP1 with $100 \, \text{fb}^{-1}$, BP2 with $3000 \, \text{fb}^{-1}$ and BP3 with $3000 \, \text{fb}^{-1}$ luminosity.}
 \label{table:cutflow}
\end{table}

\subsection{BP3 : \boldmath{$m_{xu_3} = 2$} TeV with $ 3000\, \text{fb}^{-1}$}
Finally for the last benchmark, we choose a very heavy mass of 2 TeV for the VLQ. Quite clearly the event rates would suffer 
from the very small production cross section and if we require two isolated $\tau$-jets 
then the final events yield becomes extremely low even with an integrated luminosity of 3000 fb$^{-1}$. To counter the suppression 
due to small production cross section 
we modify our signal choice to a more inclusive 
channel given by : $\geq 3$ non $\tau$-tagged jets out of which one is $b$-tagged, at least one $\tau$-tagged jet and at least one lepton in the final state. 

As we go higher in the mass of $xu_3$ the probability for the jets and the tau leptons for the signal to have higher $p_T$ values is more compared
to the backgrounds. 
Hence for BP3 with $m_{xu_3} = 2$ TeV to get a large statistics for the  $t\bar{t} + jets$ background, 
we have generated $p p \rightarrow t\bar{t} + 2j$ events exclusively at the parton level for 13 TeV LHC with the following criteria :
\begin{itemize}
 \item For each event at least one top quark decays leptonically, because at the analysis level we have considered events with at least one lepton in the final state.
\item All the jets and leptons satisfy $|\eta| < 3.0$ and the angular separation ($\Delta R$) between all pairs of final state particles are greater than 0.4 
(except for leptons where they are separated from each other with minimum angular separation 0.2).
\item All the final state objects satisfy $p_T > 10$ GeV.
 \item The two leading jets in $p_T$ satisfy $p_T(j_1) > 250$ GeV and $p_T(j_2) > 100$ GeV.
\end{itemize}
With the above cuts the parton level cross section at the leading order accuracy for $p p \rightarrow t\bar{t} + 2j$ with 
13 TeV LHC is around 6.18 pb. The same events and cross sections as in case of BP1 has been used for the other two backgrounds $t\bar{t}l^{+}l^{-} + \text{jets}$ and  
$t\bar{t}l \nu_l + \text{jets}$. 
Note that with much stronger threshold requirements for the final state jets, we expect that the lesser-order processes involving 
$t\bar{t}$ and $t\bar{t}+1j$ would not contribute much, where the extra jet comes from the showering. We then follow the usual 
procedure of using the {\tt Pythia} showering and {\tt DELPHES 3} simulation to generate the final objects from the  $p p \rightarrow t\bar{t} + 2j$
process.

\begin{figure}
 \includegraphics[width=0.7\linewidth,height=3.7in]{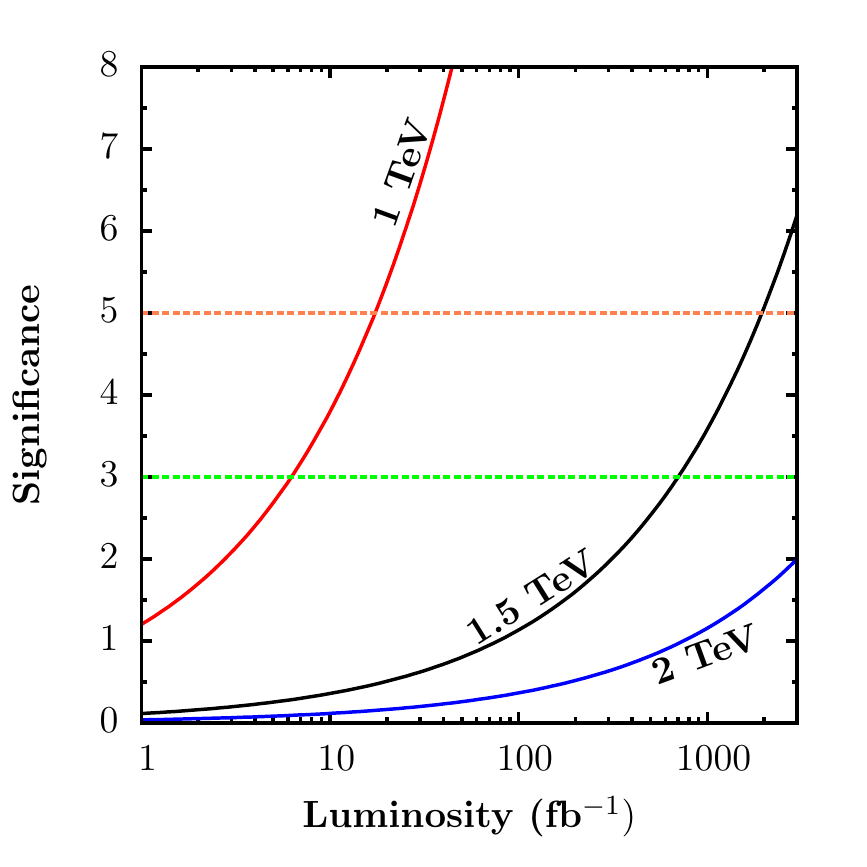}
 \caption{Significance as a function of luminosity for BP1(1 TeV), BP2(1.5 TeV) and BP3(2 TeV) for 13 TeV LHC.}
 \label{fig:significance}
\end{figure}

For our analysis, we further demand the following set of cuts on our final state events to improve the significance :
\begin{align}
 p_T(j_1) > 350 \, \text{GeV}, \, p_T(j_2) > 100 \, \text{GeV}, \, p_T(\tau_1) > 150 \, \text{GeV},  \nn \\
 \, \slashed{E_T} > 200 \, \text{GeV}, \, M_{eff} > 2.6 \, \text{TeV}. \hspace{2cm}\,
\end{align}
Here the effective mass variable ($M_{eff}$) is defined as the scalar sum of all the transverse momenta in an event and is given by 
\begin{equation}
 M_{eff} = \sum_{j \in jets} p_T(j) + \sum_{l \in leptons} p_T(l) + \slashed{E_T}.
\end{equation}
The corresponding cut flow can be seen from the table \ref{table:BP3} where we have achieved a signal significance of 
$2\sigma$ for BP3.

We plot the significance as a function of luminosity in figure \ref{fig:significance} for all the benchmark points. 
It is evident that for BP1 significance of  $5 \sigma$ can be achieved with 17.3 fb$^{-1}$ of data, hence with the already existing datasets of 36.1 fb$^{-1}$ of data would be sensitive to this mass range. With high luminosity data the BP2
can be discovered while it is only possible to exclude a 2 TeV VLQ at 2$\sigma$. In fact, an upgrade in LHC energies and 
higher luminosities would be required to access VLQ signals in such models beyond VLQ mass of $\sim 1.8$ TeV. Definitive improvements in the sensitivity is also expected with more sophisticated analysis using boosted studies for the $\tau$ 
states coming from the decay of the heavily boosted scalars.

\section{CONCLUSION}\label{section:conclusion}
In this work we have considered vector-like quarks in a leptophobic 221 model. Since SM leptons are singlets under the 
second $SU(2)$, exotic quarks are necessary to cancel triangle anomalies in this model. The exotic quarks become vector-like 
after the symmetry breaking of the full symmetry group to the SM gauge group. 
We discussed a particular mixing pattern between SM quarks and VLQs which avoids tree level FCNC interactions. 
We also find that the same mixing pattern allows for certain neutral scalars to be  flavour-conserving in nature. 
Two of these neutral scalars and their charged partner are tauphilic in nature. These scalars open up non-standard decay modes 
for the VLQs in the model. We studied the collider signatures for pair production of third generation top-like VLQ when it decays 
to final states with any of the tauphilic scalars and a third generation SM quark. Due to the mass hierarchy in the charged lepton sector, 
these scalars dominantly decay to tau leptons with more than $99\%$ probability.
We do an analysis for the signal of such VLQs,  pair produced at the LHC with $\sqrt{s} = 13$ TeV through the 
$\geq 3j(1b) \, + \geq 2\tau \,  + \geq 1l$ final state, dictated by the decay properties of the VLQ and the new tauphilic scalars. 
We use mass threshold driven kinematic selections for the final state objects and show the values of the integrated luminosity 
required for the discovery of such a top-like VLQ for different benchmark points.
We find that the amount of data collected till date by the ATLAS and CMS collaborations for 13 TeV LHC is sufficient to confirm or refute 
the existence of such a scenario for a 1 TeV top-like VLQ. Heavier VLQ masses up to 1.8 TeV would be accessible with the 
HL option of LHC. This study also highlights an important point of caution for VLQ searches in the standard decay channels 
carried out at the LHC, that any new physics scenario which may have additional gauge bosons and scalars can alter the VLQ searches in a significant way and therefore alternative channels of search should also be considered, as the VLQ mass limits 
crucially depend on them \cite{Dobrescu:2016pda}. 
\begin{acknowledgments}
K.D. would like to thank Tianjun Li for useful discussions, Jyotiranjan Beuria for help regarding SARAH and SPheno, Manuel E. Krauss and Subhadeep Mondal for help regarding SARAH.
The work was partially supported by funding available from the Department of Atomic 
Energy, Government of India, for the Regional Centre for Accelerator-based Particle Physics (RECAPP), 
Harish-Chandra Research Institute. The research of K.D. was supported in part by the INFOSYS
scholarship for senior students at the Harish-Chandra Research Institute.
The authors acknowledge the use of the High Performance Scientific Computing facility at RECAPP and HRI. 
\end{acknowledgments}
\appendix
\section{Tadpole equations}\label{appendix:tadpole}
The set of tadpole equations 
$
 \big\{\frac{\partial V}{\partial {\phi_1^0}^r} = 0, \frac{\partial V}{\partial {\phi_2^0}^r} = 0, \frac{\partial V}{\partial {\chi^0}^r} = 0, 
 \frac{\partial V}{\partial {{\chi^\prime}^0}^r} = 0\big\}
$
in terms of $\mu_1^2, \, \mu_2^2, \, \mu_3^2, \, \mu_4^2$ are given by
\begingroup
\allowdisplaybreaks[3]
\begin{align}
\mu_1^2 &= \frac{1}{2(v_1^2-v_2^2)}\big\{\alpha _1 u^2 v_1^2-\alpha _1 u^2 v_2^2-\alpha _3 u^2 v_2^2 +2 \lambda _1 v_1^4+4 \lambda _4 v_2 v_1^3-4 \lambda _4 v_2^3 v_1-2 \lambda _1 v_2^4 \nn \\
  & \hspace{5cm}+ \sqrt{2} u v_3 \left(M_1 v_1-M_2 v_2\right)+ v_3^2 \left(\beta _1 v_1^2-\left(\beta _1+\beta_3\right) v_2^2\right)\big\}, \\
\mu_2^2 &= \frac{1}{4  \left(v_1^2-v_2^2\right)} \big\{ 2 \alpha _2 u^2 v_1^2  +\alpha _3 u^2 v_2 v_1-2 \alpha _2 u^2 v_2^2 +2 \lambda _4 v_1^4+8 \lambda _2 v_2 v_1^3 + 4 \lambda _3 v_2 v_1^3 \nn \\
  & \hspace{4cm}-8 \lambda_2 v_2^3 v_1-4 \lambda _3 v_2^3 v_1-2 \lambda _4 v_2^4 + \sqrt{2} u v_3 \left(M_2 v_1-M_1 v_2\right) \nn \\
  & \hspace{9cm}+v_3^2 \left(2 \beta _2 \left(v_1^2-v_2^2\right)+\beta _3 v_1 v_2\right) \big\},\\
\mu_3^2 &=\frac{1}{2 v_3} \big\{\sqrt{2} u \left(M_1 v_1+M_2 v_2\right)+v_3 \left(\rho _3 u^2+\beta _1 v_1^2+4 \beta _2 v_1 v_2+\left(\beta_1+\beta _3\right) v_2^2\right)+2 \rho _1 v_3^3\big\}, \\
\mu_4^2 & = \frac{1}{2 u} \big\{  v_1 \left(\sqrt{2} M_1 v_3+4 \alpha _2 u v_2\right)+\sqrt{2} M_2 v_2 v_3+u \left(2 \rho _2 u^2+\left(\alpha_1+\alpha _3\right) v_2^2+\rho _3 v_3^2\right) \nn \\
    & \hspace{12cm}+\alpha _1 u v_1^2  \big\}.
\end{align}
\section{Mass matrices for the scalar Sector}\label{appendix:scalar_massmatrices}
\subsection{CP even scalars}\label{appendix:cp_even_mass_matrices}
The components of the CP even scalar sector mass square matrix($M_S^2$) in the $({\phi_1^0}^r,\, {\phi_2^0}^r, \, {{\chi^\prime}^0}^r, \, {{\chi^0}^r})$ basis is given by 
\begin{align}
 (M^2_S)_{11} &= \frac{1}{2 \left(v_1^2-v_2^2\right)}\big\{v_2 \left(\sqrt{2} M_2 u v_3+\alpha _3 u^2 v_2-4 \left(2 \lambda _2+\lambda _3\right) v_2^3\right)-v_1 \left(\sqrt{2} M_1 u v_3+8 \lambda _4 v_2^3\right) \nn \\
               &\hspace{5cm}  +\beta _3 v_2^2 v_3^2+4 \lambda _1 v_1^4+8 \lambda _4 v_2 v_1^3+4\left(-\lambda _1+2 \lambda _2+\lambda _3\right) v_2^2 v_1^2\big\} \nn \\
 (M^2_S)_{12} &=  \frac{1}{2 \left(v_1^2-v_2^2\right)} \big\{-v_1 \left(\sqrt{2} M_2 u v_3+\alpha _3 u^2 v_2+4 \left(\lambda _1+2 \lambda _2+\lambda _3\right) v_2^3\right) +\sqrt{2} M_1 u v_2 v_3 \nn \\
               &  \hspace{5cm}  -\beta _3 v_2 v_3^2 v_1+4 \lambda _4 v_1^4+4 \left(\lambda _1+2 \lambda_2+\lambda _3\right) v_2 v_1^3-4 \lambda _4 v_2^4\big\} \nn \\
 (M^2_S)_{13} &= \frac{M_1 v_3}{\sqrt{2}}+u \left(\alpha _1 v_1+2 \alpha _2 v_2\right) \nn \\
 (M^2_S)_{14} &= \frac{M_1 u}{\sqrt{2}}+v_3 \left(\beta _1 v_1+2 \beta _2 v_2\right) \nn \\
  (M^2_S)_{22} &= \frac {1}{2 \left(v_1^2-v_2^2\right)} \big\{-v_1 \left(\sqrt{2} M_1 u v_3+8 \lambda _4 v_2^3\right)+\sqrt{2} M_2 u v_2 v_3 +\beta _3 v_3^2 v_1^2    +4 \left(2 \lambda _2+\lambda_3\right) v_1^4   \nn \\
                 & \hspace{5cm}  +8 \lambda _4 v_2 v_1^3-4 \lambda _1 v_2^4 + v_1^2 \left(\alpha _3 u^2 + 4\left(\lambda _1-2 \lambda _2-\lambda _3\right) v_2^2\right)\big\} \nn \\
 (M^2_S)_{23} &= \frac{M_2 v_3}{\sqrt{2}}+u \left(2 \alpha _2 v_1+\left(\alpha _1+\alpha _3\right) v_2\right) \nn \\
 (M^2_S)_{24} &=  \frac{M_2 u}{\sqrt{2}}+v_3 \left(2 \beta _2 v_1+\left(\beta _1+\beta _3\right) v_2\right) \nn \\
 (M^2_S)_{33} &= \frac{1}{2 u}\big\{4 \rho _2 u^3-\sqrt{2} v_3 \left(M_1 v_1+M_2 v_2\right)\big\} \nn \\
 (M^2_S)_{34} &= \frac{M_1 v_1}{\sqrt{2}}+\frac{M_2 v_2}{\sqrt{2}}+\rho _3 \, u \, v_3 \nn \\
 (M^2_S)_{44} &=  \frac{1}{2 v_3}\big\{4 \rho _1 v_3^3-\sqrt{2} u \left(M_1 v_1+M_2 v_2\right)\big\}
\end{align}
\subsection{CP odd scalars}\label{appendix:cp_odd_mass_matrices}
The components of the CP odd scalar sector mass square matrix($M_P^2$) in the $({\phi_1^0}^i,\, {\phi_2^0}^i, \, {{\chi^\prime}^0}^i, \, {{\chi^0}^i})$ basis is given by
\begin{align}
 (M^2_P)_{11} &=  \frac{1}{2 \left(v_1^2-v_2^2\right)}\big\{\sqrt{2} u v_3 \left(M_2 v_2-M_1 v_1\right)+v_2^2 \left(\alpha _3 u^2-4 \left(2 \lambda _2-\lambda _3\right)\left(v_1^2-v_2^2\right)\right)+\beta _3 v_3^2 v_2^2\big\} \nn \\
 (M^2_P)_{12} &=  \frac{1}{2 \left(v_1^2-v_2^2\right)}\big\{\sqrt{2} u v_3 \left(M_2 v_1-M_1 v_2\right)+v_1 v_2 \left(\alpha _3 u^2-4 \left(2 \lambda _2-\lambda _3\right)\left(v_1^2-v_2^2\right)\right)+\beta _3 v_1 v_2 v_3^2\big\} \nn \\
 (M^2_P)_{13} &= -\frac{M_1 v_3}{\sqrt{2}} \nn \\
 (M^2_P)_{14} &= \frac{M_1 u}{\sqrt{2}} \nn \\
  (M^2_P)_{22} &= \frac{1}{2 \left(v_1^2-v_2^2\right)}\big\{\sqrt{2} u v_3 \left(M_2 v_2-M_1 v_1\right)+v_1^2 \left(\alpha _3 u^2-4 \left(2 \lambda _2-\lambda _3\right)   \left(v_1^2-v_2^2\right)\right)+\beta _3 v_3^2 v_1^2\big\} \nn \\
  (M^2_P)_{23} &=  \frac{M_2 v_3}{\sqrt{2}} \nn \\
  (M^2_P)_{24} &= -\frac{M_2 u}{\sqrt{2}} \nn \\
  (M^2_P)_{33} &= -\frac{v_3 \left(M_1 v_1+M_2 v_2\right)}{\sqrt{2} u} \nn \\
  (M^2_P)_{34} &= \frac{M_1 v_1+M_2 v_2}{\sqrt{2}} \nn \\
  (M^2_P)_{44} &=-\frac{u \left(M_1 v_1+M_2 v_2\right)}{\sqrt{2} v_3}
 \end{align}
 \subsection{Charged scalars}\label{appendix:charged_scalar_mass_matrices}
The mass square matrix for the charged scalars ($M_C^2$) in the $(\phi_1^+, \, \phi_2^+, \, {\chi^\prime}^+, \, {\chi^+})$ basis is given by
\begin{align}
 (M^2_C)_{11} &= \frac{1}{2 \left(v_1^2-v_2^2\right)}\big\{u \left(\sqrt{2} v_3 \left(M_2 v_2-M_1 v_1\right)+\alpha _3 u   v_2^2\right)+\beta _3 v_1^2 v_3^2\big\} \hspace{4.5cm}\nn \\
 (M^2_C)_{12} &= \frac{1}{2 \left(v_1^2-v_2^2\right)}\big\{u \left(\sqrt{2} v_3 \left(M_2 v_1-M_1 v_2\right)+\alpha _3 u v_1 v_2\right)+\beta _3 v_1 v_2 v_3^2\big\} \nn \\
 (M^2_C)_{13} &= -\frac{M_2 v_3}{\sqrt{2}}-\frac{1}{2} \alpha _3 \, u \, v_2 \nn \\
 (M^2_C)_{14} &= \frac{M_1 u}{\sqrt{2}}-\frac{1}{2} \beta _3 v_1 v_3   \nn \\
 (M^2_C)_{22} &= \frac{1}{2 \left(v_1^2-v_2^2\right)}\big\{u \left(\sqrt{2} v_3 \left(M_2 v_2-M_1 v_1\right)+\alpha _3 u v_1^2\right)+\beta _3 v_2^2 v_3^2\big\} \nn \\
 (M^2_C)_{23} &= \frac{M_1 v_3}{\sqrt{2}}-\frac{1}{2} \alpha _3 \, u \, v_1 \nn \\
 (M^2_C)_{24} &= -\frac{M_2 u}{\sqrt{2}}-\frac{1}{2} \beta _3 v_2 v_3 \nn \\
 (M^2_C)_{33} &= \frac{1}{2} \left(\alpha _3 \left(v_1^2-v_2^2\right)-\frac{\sqrt{2} v_3 \left(M_1 v_1+M_2 v_2\right)}{u}\right) \nn \\
 (M^2_C)_{34} &= \frac{M_2 v_1+M_1 v_2}{\sqrt{2}} \nn \\
 (M^2_C)_{44} &= \frac{1}{2 v_3}\big\{\beta _3 \left(v_1^2-v_2^2\right) v_3-\sqrt{2} u \left(M_1 v_1+M_2   v_2\right)\big\}
\end{align}
\endgroup

\bibliographystyle{JHEP}
\bibliography{ref.bib}

\end{document}